\documentclass[times,twocolumn,final]{elsarticle}

\usepackage{framed,multirow}
\usepackage{amssymb}
\usepackage{amsmath}
\usepackage{latexsym}
\usepackage{url}
\usepackage{xcolor}

\usepackage{balance}
\usepackage{hyperref}
\usepackage[vertfit]{breakurl}
\usepackage{graphicx}
\usepackage{caption}
\usepackage{subcaption}

\usepackage{threeparttable}
\usepackage{booktabs}
\usepackage{gensymb}

\usepackage{float}
\usepackage{dblfloatfix}

\usepackage[switch]{lineno}
\usepackage[normalem]{ulem}

\newcommand{\ie}{\textit{i}.\textit{e}.\,}
\newcommand{\eg}{\textit{e}.\textit{g}.\,}
\newcommand{\etal}{\textit{et~al.\,}}

\begin{document}

\begin{frontmatter}

	\title{Combining unsupervised and supervised learning for predicting the final stroke lesion}%
	\author[UM-DEI,UM-DI]{Adriano Pinto \corref{main}}
	\cortext[main]{Corresponding author: Department of Industrial Electronics, Campus Azurém, Guimarães, Portugal.}
	\ead{id6376@alunos.uminho.pt}
	
	\author[UM-DEI,UM-DI]{S\'ergio Pereira}
	\author[INSEL-SCAN]{Raphael Meier}
	\author[INSEL-SCAN]{Roland Wiest}
	\author[UM-DI]{Victor Alves}
	\author[INSEL-AI]{Mauricio Reyes}
	\author[UM-DEI]{Carlos A. Silva}
	\ead{csilva@dei.uminho.pt}

	\address[UM-DEI]{Center MEMS of University of Minho, Campus of Azur\'em, 4800-058 Guimar\~aes Portugal}
	\address[UM-DI]{Center Algoritmi, University of Minho, Braga, Portugal}
	\address[INSEL-AI]{Healthcare Imaging A.I., Insel Data Science Center, Bern University Hospital, Switzerland}
	\address[INSEL-SCAN]{Support Center for Advanced Neuroimaging, University Institute for Diagnostic and Interventional Neuroradiology, Bern University Hospital, Switzerland}

\begin{abstract}
	\small{	Predicting the final ischaemic stroke lesion provides crucial information regarding the volume of salvageable hypoperfused tissue, which helps physicians in the difficult decision-making process of treatment planning and intervention. Treatment selection is influenced by clinical diagnosis, which requires delineating the stroke lesion, as well as characterising cerebral blood flow dynamics using neuroimaging acquisitions. Nonetheless, predicting the final stroke lesion is an intricate task, due to the variability in lesion size, shape, location and the underlying cerebral haemodynamic processes that occur after the ischaemic stroke takes place. Moreover, since elapsed time between stroke and treatment is related to the loss of brain tissue, assessing and predicting the final stroke lesion needs to be performed in a short period of time, which makes the task even more complex. Therefore, there is a need for automatic methods that predict the final stroke lesion and support physicians in the treatment decision process. We propose a fully automatic deep learning method based on unsupervised and supervised learning to predict the final stroke lesion after 90 days. Our aim is to predict the final stroke lesion location and extent, taking into account the underlying cerebral blood flow dynamics that can influence the prediction. To achieve this, we propose a two-branch Restricted Boltzmann Machine, which provides specialized data-driven features from different sets of standard parametric Magnetic Resonance Imaging maps. These data-driven feature maps are then combined with the parametric Magnetic Resonance Imaging maps, and fed to a Convolutional and Recurrent Neural Network architecture. We evaluated our proposal on the publicly available ISLES 2017 testing dataset, reaching a Dice score of 0.38, Hausdorff Distance of 29.21 mm, and Average Symmetric Surface Distance of 5.52 mm.}

\end{abstract}

\begin{keyword} 
Deep Learning\sep Image Prediction\sep 
Magnetic Resonance Imaging\sep Stroke 
\end{keyword}

\end{frontmatter}


\graphicspath{{images/}}

\section{Introduction}
\label{sec:intro}
	
	Stroke is the second leading cause of death worldwide \citep{world2014global}, being classified in two types: ischaemic and haemorrhagic \citep{grysiewicz2008epidemiology}. Ischaemic stroke is the most common type, resulting from an occlusion of a vessel, which can be caused by thrombolysis, haemodynamic factors, or embolic causes \citep{grysiewicz2008epidemiology}. Due to vessel occlusion, the insufficient supply of oxygenated blood to brain cells leads to hypoperfused brain tissue, triggering cellular mechanisms to preserve the integrity of the cell. The hypoperfused area consists of tissue at risk that can be salvaged, being designated penumbra. As time passes, in the absence of flow restoration or sufficient collateral blood flow supply, the hypoperfused tissue eventually reaches a non-salvageable state designated core or infarct tissue \citep{memezawa1992penumbral}.

	Diagnosis and treatment of ischaemic stroke relies on neuroimaging acquisitions, where Computed Tomography (CT) and Magnetic Resonance Imaging (MRI) are the preferred imaging modalities \citep{gonzalez2007acute}. CT imaging remains the most used acquisition due to its rapidity and availability \citep{gonzalez2007acute}. However, multi-parametric MRI provides a higher sensitivity in detecting early ischaemic stroke and assessing the penumbra region \citep{gonzalez2007acute}. Treatment consists in restoring tissue perfusion levels, also known as reperfusion, by performing mechanical thrombectomy or thrombolysis. Since ischaemic stroke is a dynamic process that evolves over time, the treatment is only possible up to 24 hours, where viable neurones still persist \citep{el2017thrombolysis,zivelonghi2018mechanical}. So, expert physicians must evaluate the benefits and risks of mechanical thrombectomy before an intervention, since it may cause haemorrhage, vascular injury, and other complications \citep{powers20182018}. If performed, the success of the intervention is assessed radiologically via angiography imaging and scored by a qualitative expert-generated scale designated the standardized Thrombolysis in Cerebral Infarction (TICI) scale \citep{higashida2003trial}. During the decision-making process, the physician needs to assess the nature and location of the lesion alongside pathophysiological factors such as age, presence of comorbidities, and collateral circulation \citep{liebeskind2003collateral}. The latter is of utmost importance in ischaemic stroke. The presence of collateral circulation, where a secondary network of vessels is responsible for granting cerebral blood flow to the lesioned tissue, increases the chances of a successful reperfusion \citep{liebeskind2003collateral}. Asserting the potential efficacy of treatment can be time-consuming and prone to inter- and intra-variability among physicians, which is further potentiated when performed in a clinical emergency environment \citep{coutts2003reliability}. Moreover, since time is critical, MRI acquisitions are optimized for speed, which is accomplished by reducing the resolution \citep{gonzalez2011acute}, making the prediction of the final stroke lesion an intricate task. Thus, automatic prediction of a stroke lesion at a given time since stroke has a great potential to guide physicians in this time-critical decision-making process.

	\begin{figure*}[!ht]
	\centering
    \begin{subfigure}[h]{0.45\textwidth}
        \includegraphics[width=\textwidth]{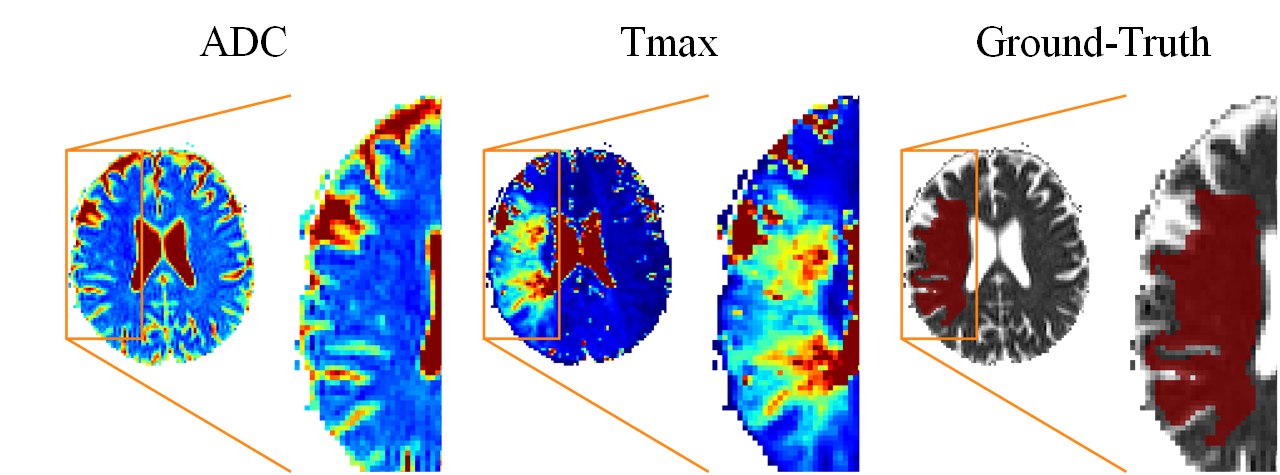}
        \caption{} \label{fig:val_example_case_1}
    \end{subfigure}
    ~
    \begin{subfigure}[h]{0.45\textwidth}
        \includegraphics[width=\textwidth]{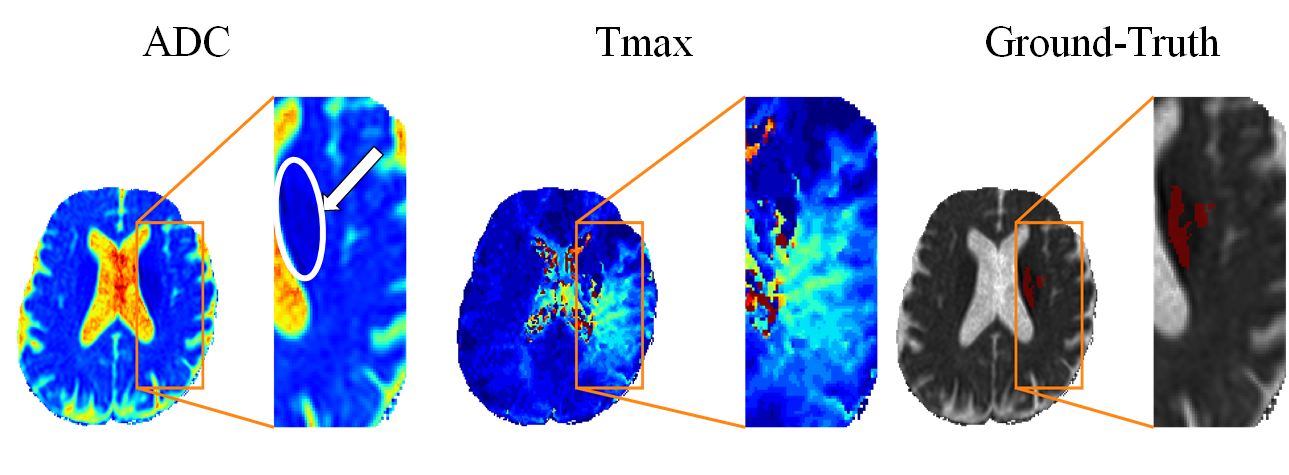}
        \caption{} \label{fig:val_example_case_2}
    \end{subfigure}
    \caption{ADC and Tmax parametric maps of two patient cases from ISLES 2017 training set, and the final lesion delineated at a 90-day follow-up, overlapped with the onset ADC: patient 0036 (Figure \ref{fig:val_example_case_1}) with an unsuccessful reperfusion, and patient 0006 (Figure \ref{fig:val_example_case_2}), where the clinical intervention was successful.} 
    \label{fig:val_example_cases}
\end{figure*}

	We propose a novel automatic method based on unsupervised and supervised deep learning. We utilize Restricted Boltzmann Machines (RBMs) to jointly characterise the lesion and blood flow information through a two-pathway architecture, trained with two subsets of standard parametric MRI maps. One subset encompasses the Time-To-Peak (TTP), Mean Transit Time (MTT), Time-to-Maximum (Tmax), and Apparent Diffusion Coefficient (ADC). The second set contains the ADC, the relative Cerebral Blood Volume (rCBV), and the relative Cerebral Blood Flow (rCBF). In a second stage, the feature maps computed by the RBMs are combined with the standard parametric MRI maps to form the input of a supervised deep learning architecture composed by Convolutional Neural Networks (CNNs) and Recurrent Neural Networks (RNNs). The proposed architecture was evaluated using the publicly-available ISLES 2017 dataset.

\subsection{On the complexity of predicting the final infarct stroke lesion}
			
	In acute ischaemic stroke, the clinical evaluation of the standard parametric maps (\eg ADC and Tmax) can identify infarct tissue and tissue that will infarct in the absence of therapeutic intervention. In this analysis, the infarct tissue, is identified by the hypointense regions of the ADC map, which characterize tissue with limited diffusion \citep{butcher2010acute1}. Hypoperfused tissue, \ie tissue that will infarct, is identified by hyperintense regions of the Tmax map, indicating an increased arrival time of contrast agent \citep{butcher2010acute2}. However, to correctly predict the final ischaemic stroke lesion, besides considering the complex time-evolving transformation of hypoperfused tissue to infarcted tissue, it is also necessary to appraise the impact of the clinical intervention, thrombectomy, on the underlying brain perfusion and diffusion.

	A successful thrombectomy should restore the perfusion levels, recovering the hypoperfused tissue. However, several factors may affect the reperfusion, limiting the degree of success of the intervention. To better understand the nuances of the clinical intervention, consider the cases presented in Figure \ref{fig:val_example_cases}. In the first case, Figure \ref{fig:val_example_case_1}, the ADC does not present any hypointense region, so no infarct tissue may be identified, and we should expect a complete recovery of the hypoperfused tissue indicated by Tmax; however, the follow-up delineation obtained after thrombectomy presents a large final lesion, which is explained by an unsuccessful intervention. In the second case, Figure \ref{fig:val_example_case_2}, we observe a final infarct lesion that is smaller than the hypointense region present in the ADC (Figure \ref{fig:val_example_case_2} arrow). This indicates reversible diffusion restriction, which is a rare case \citep{labeyrie2012diffusion} and was only possible to identify by a follow-up T2-weighted acquisition. So, an automatic method for predicting the final stroke lesion has not only to capture the time-evolving process of diffusion and perfusion, but also to consider directly or indirectly the degree of success of the thrombectomy, which may condition the final lesion either to be confined to the hypointense region of the ADC map, or to grow to brain tissue areas that are hyperintense in the Tmax. Due to the time-evolving process of diffusion and perfusion, the complexity of predicting the lesion will aggravate as we move from a target window of some days to several months.

	The complexity of the evaluation process may be also observed in the inter-rater agreement of expert radiologists in ISLES 2017 dataset, which obtained a Dice score of $0.58 \pm 0.20$ on delineating the lesion using a 90-day follow-up T2-weighted acquisition \citep{winzeck2018isles}.

	\subsection{Previous Work}

		Contrary to stroke lesion segmentation, where several methods have already been proposed \citep{rekik2012medical,maier2017isles}, the complexity of predicting the final stroke lesion has only recently attracted attention in the medical imaging community. For predicting the final stroke lesion several methods have been already proposed based on multivariate linear regression models \citep{scalzo2012,rose2001mri,kemmling2015}, decision trees \citep{mckinley,bauer2014towards}, and CNNs \citep{choi2016ensemble}. Furthermore, with the release of Ischaemic Stroke LEsion Segmentation (ISLES) Challenge in 2016 and 2017, new methods have been proposed. These aim to predict at a 90-day time-window.

		\citet{rose2001mri} proposed a two-stage method based on parametric perfusion and diffusion MRI maps. On the first stage, the method defines a region of interest (ROI) based on the intensity signal of the standard parametric maps, the MTT, Cerebral Blood Flow (CBF), Cerebral Blood Volume (CBV), and Diffusion-Weighted Imaging (DWI). The second stage performs stroke tissue prediction using Gaussian mixture models trained in different sets of parametric maps. \citet{bauer2014towards} used Random Forests to segment or predict the final stroke lesion depending on whether acute stroke imaging or three-month follow-up imaging was available, respectively. \citet{mckinley} also used a two-stage classification approach as in \citet{rose2001mri} for lesion characterisation and lesion prediction, where each stage consists of two sets of Random Forests (RFs) classifiers. The first stage aims to define a ROI that encompasses the hypoperfused region. In the first set, each classifier is trained with features extracted from different sets of MRI parametric maps. Having defined the location and extension of the lesion, a second set of two RFs performs stroke tissue prediction. Such classifiers were trained on different sets of patients, stratified by the TICI score. One classifier is trained in patients with unsuccessful reperfusion interventions, whereas a second classifier is trained in patients with successful reperfusion. The final prediction is obtained by combining the results of both classifiers, using a logistic regression model. \citet{scalzo2012} proposed a framework for stroke tissue prediction, which characterises the state of the lesion four days after clinical intervention (thrombectomy). From the Fluid Attenuation Inversion Recovery (FLAIR) MRI sequence, ADC and Tmax MRI maps, the method applies a regression model that learns the behaviour of neighbouring voxels within a cuboid. \citet{kemmling2015} proposed a multi-modality approach based on CT and MRI maps with non-imaging clinical meta-data, namely the TICI score and the time to treatment of each patient, to perform stroke tissue prediction.

		In another line of research, authors have investigated the use of deep learning \citep{choi2016ensemble,nielsen2018prediction,robben2020prediction} for stroke tissue prediction. \citet{choi2016ensemble}, the winner approach at ISLES 2016 Challenge, proposed an ensemble of twelve CNN architectures, grouped into two sets of networks. The first group comprehends four 3D U-Nets \citep{ronneberger2015u} performing voxel-wise tissue prediction. The second group of networks uses two-pathway Fully Connected Networks (FCNs) performing two types of patch-wise classification. One set of FCNs classifies a patch as lesion if it includes any lesion voxel. The other set of FCNs classifies a patch as lesion if the central voxel is a lesion. After merging the two pathway FCN, the method incorporates meta-data by adding a dense layer of clinical predictors merged with the imaging output of each network. The final stroke lesion prediction results from a weighted merging of all models. \citet{mokdeep} applied deep adversarial training for stroke tissue prediction in an ensemble of U-Nets. \citet{monteirofully} proposed a method based on the V-Net architecture \citep{milletari2016v}. The training was conducted with a custom loss function that applies a weighted sum between Dice score and cross entropy. \citet{lucas20172d} proposed the use of a U-Net architecture, which combines patches from the MRI maps in the same slice, with patches from 3 neighbouring slices and 2 hemispheric flips. In the expanding path of the U-Net, each level computes a Dice loss for the healthy tissue and for the final lesion, after the softmax activation. Afterwards, all losses are summed up, having the loss of the lesion and healthy tissue weighted according to a prior probability \citep{winzeck2018isles}. \citet{robben2017dual} employed a CNN-based architecture inspired by \citet{kamnitsas2017efficient}. The authors proposed to combine the MRI inputs with clinical meta-data, before feeding them to each branch of a two-pathway 3D network. In the first branch the input is kept with the original resolution, while in the second branch the input resolution was lowered by a factor of 3. The output of each branch is transformed to the same scale and merged by two fully connected layers. The network is trained with four different sets of hyper-parameters. These four networks are used as an ensemble, whose prediction is obtained by averaging the output of each one. Similarly, \citet{niu2018improved} used multiple scales of overlapping 3D patches to capture local and global spatial information. In the review paper of \citet{winzeck2018isles}, Rivera \etal also built on the work of \citet{kamnitsas2017efficient} and \citet{milletari2016v}, by proposing a scheme to extract different patch resolutions, independent of each other, that are fed into four different paths. Afterwards, a fully connected layer combines all the outputs to perform stroke tissue prediction. \citet{pisovneural} employed an ensemble strategy by combining different CNN-based architectures to overcome the strong anisotropy of the data. As summarized by \citet{winzeck2018isles}, the work of Yoon \etal proposed a two-stage gated CNN. In a first stage, the authors perform lesion detection and delineation. Afterwards, based on the probability maps of the first stage, a second CNN architecture processes the regions where the probability maps of healthy tissue and lesion are close to each other. \citet{pinto2018enhancing} made use of temporal perfusion imaging, the Dynamic Susceptibility Contrast-MRI, in a U-Net architecture. This architecture aims to temporally process and extract deep features, which are then combined with a second feature step of another U-Net network, which was trained on the standard parametric maps. Using a large CT dataset, \citet{robben2020prediction} predicted the final infarct stroke lesion with a temporal window ranging from $24$ hours to 5 days. The authors considered spatio-temporal CT perfusion as input to a deep neural network inspired in the architecture proposed by \citet{kamnitsas2017efficient}. Additionally, the model combines CT neuroimaging with clinical meta-data. \citet{nielsen2018prediction} proposed a method based on the SegNet architecture \citep{badrinarayanan2015segnet}, predicting on a 30-day follow-up acquisition based on a private dataset. 

		Principal and collateral blood flow has been considered either directly by modelling the temporal perfusion imaging \citep{pinto2018enhancing}, or indirectly by perfusion and diffusion parametric maps \citep{choi2016ensemble, maier2017isles, scalzo2012}, or through clinical information that characterises the success of the revascularization \citep{mckinley}. We hypothesize that modelling the haemodynamics of the brain when artery occlusion occurs can be beneficial for predicting the final stroke lesion. So, in this work, we investigate the representation of the haemodynamics through an unsupervised learning model. Contrary to previous approaches, we propose grouping the input maps according to their subjacent physical meaning and encoding each group separately with an RBM. As groups, we investigated the time-resolved perfusion maps (Tmax, TTP, MTT), and the blood-flow-dynamic related maps (rCBF, rCBV) \citep{butcher2010acute1,butcher2010acute2}. Our proposal of combining features obtained unsupervisedly and supervisedly was motivated by the knowledge that unsupervised models learn structural features of the original image, while the supervised models learn features conditioned on the label, so there is potential for obtaining richer and more discriminative features by joining both types of models.		

			\begin{figure*}[!ht]
			\includegraphics[width=\textwidth]{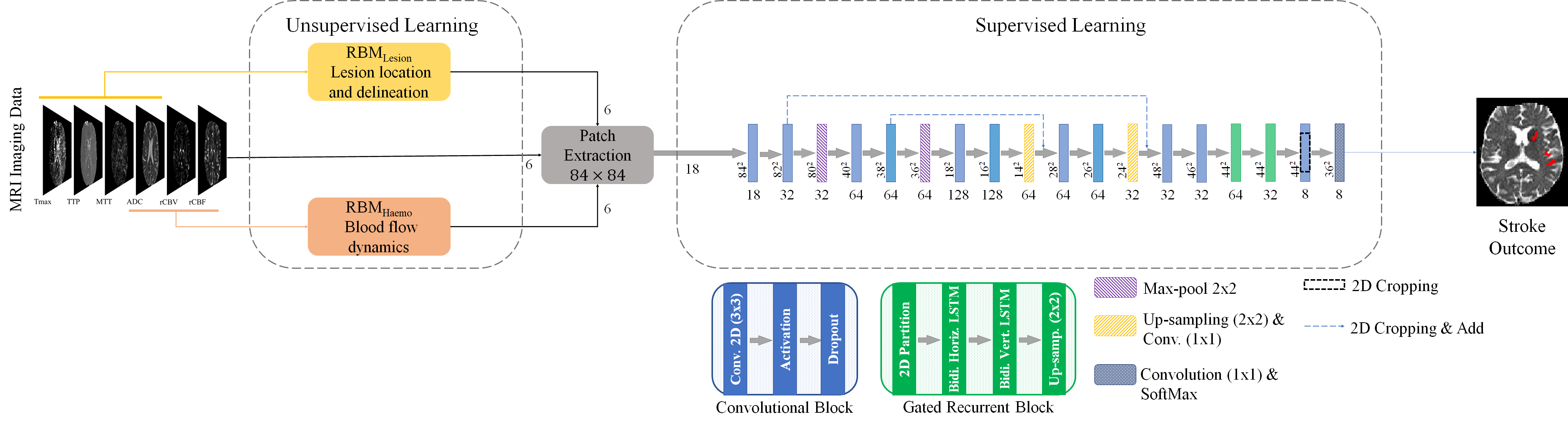}
			\centering
			\caption{Overview of the proposed method for predicting the final stroke lesion. In the supervised learning block, the input data dimensions are defined for each operation.}
			\label{fig:architecture}
	\end{figure*}
	\subsection{Contributions}

		This work presents an automatic approach for predicting the final stroke lesion, using onset neuroimaging data. The main contributions are:

		\begin{itemize}           
			\item[-] The use of unsupervised models for extracting structural features of time-resolved perfusion and blood-flow-dynamic related MRI maps for predicting stroke lesion.

			\item[-] The use of local and long spatial context provided by gated recurrent neural networks for relating structural features and image information when learning features conditioned on the label in a supervised model.
			
			\item[-] The proposal of a competitive system which outperforms state-of-the-art methods to predict the final infarct stroke lesion, in ISLES 2017 Challenge dataset.
			
	   \end{itemize}

		The remainder of the paper is organized as follows. Section \ref{sec:method} describes the fundamental components of the proposed method. Section \ref{sec:eval} describes the dataset, the evaluation procedure and the setup. The results and the discussion are addressed in Section \ref{sec:results}. Finally, in Section \ref{sec:conclusion} we present the main conclusions.


\section{Methods}
\label{sec:method}
	
	In this work, predicting the final infarct stroke lesion consists of delineating the lesion's spatial extent at a 90-day follow-up time-point, using multi-parametric MRI imaging, namely the ADC, MTT, TTP, Tmax, rCBF, and rCBV, which are acquired at the onset time-point. The architecture of the proposed system and its main components are described in the following subsections.
	
	\subsection{Architecture}
		
		The overall architecture of the proposed method can be divided into two functional blocks, as shown in Figure \ref{fig:architecture}.
		
		The first functional block performs unsupervised representation learning using two unsupervised models, namely RBMs. This unsupervised block provides new features that represent structural information that complements the standard parametric MRI maps, enhancing the capacity of our model to predict the final infarct lesion. In our approach, we aim to model the clinical procedure, which first locates and delineates the lesion at current time, and then considers the blood flow haemodynamic that might influence the final stroke lesion prediction. This procedure is encoded in our two-path RBM. The first RBM is responsible for capturing information on lesion location and extension, referred to as the RBM\textsubscript{Lesion}. The second RBM, RBM\textsubscript{Haemo}, aims to capture blood flow haemodynamics information (\eg collateral circulation), which has been identified as a key factor by physicians when assessing stroke final infarct lesion in clinical reports \citep{berkhemer2016collateral,menon2015differential}. On one hand, to locate the onset ischaemic stroke lesion, the RBM\textsubscript{Lesion} considers standard parametric maps that characterise the arrival times and mean transit times of the contrast agent. In the presence of an ischaemic lesion, the occluded vessel can decrease or interrupt the normal brain perfusion, translating into hyperintense regions on time-related parametric maps \citep{butcher2010acute2}. On the other hand, the RBM\textsubscript{Haemo} considers standard parametric maps that characterise the amount of blood being delivered in unit of time, which correlates to the cerebral blood flow haemodynamics \citep{butcher2010acute2}. Thus, the RBM\textsubscript{Lesion} considers the MTT, TTP and Tmax perfusion maps, while the RBM\textsubscript{Haemo} the rCBV and rCBF perfusion maps. Regarding the ADC standard diffusion map, it is present in both RBM\textsubscript{Lesion} and RBM\textsubscript{Haemo}, since it provides higher brain structural information and allows the identification of tissue that is already infarcted. This separation of the input imaging allows the RBM to learn specific feature sets, which may enable the method to analyse difficult cases where information concerning the blood flow can have a favourable impact on the lesion prediction.
	
		The second functional block consists of a deep learning architecture that comprehends 2D convolutional blocks in a U-Net-based structure, alongside recurrent blocks. As imaging input data, we combine the standard parametric maps with feature maps from each RBM, totalling $18$ input feature maps.
	
		\begin{figure*}[!ht]
			\includegraphics[width=0.7\textwidth]{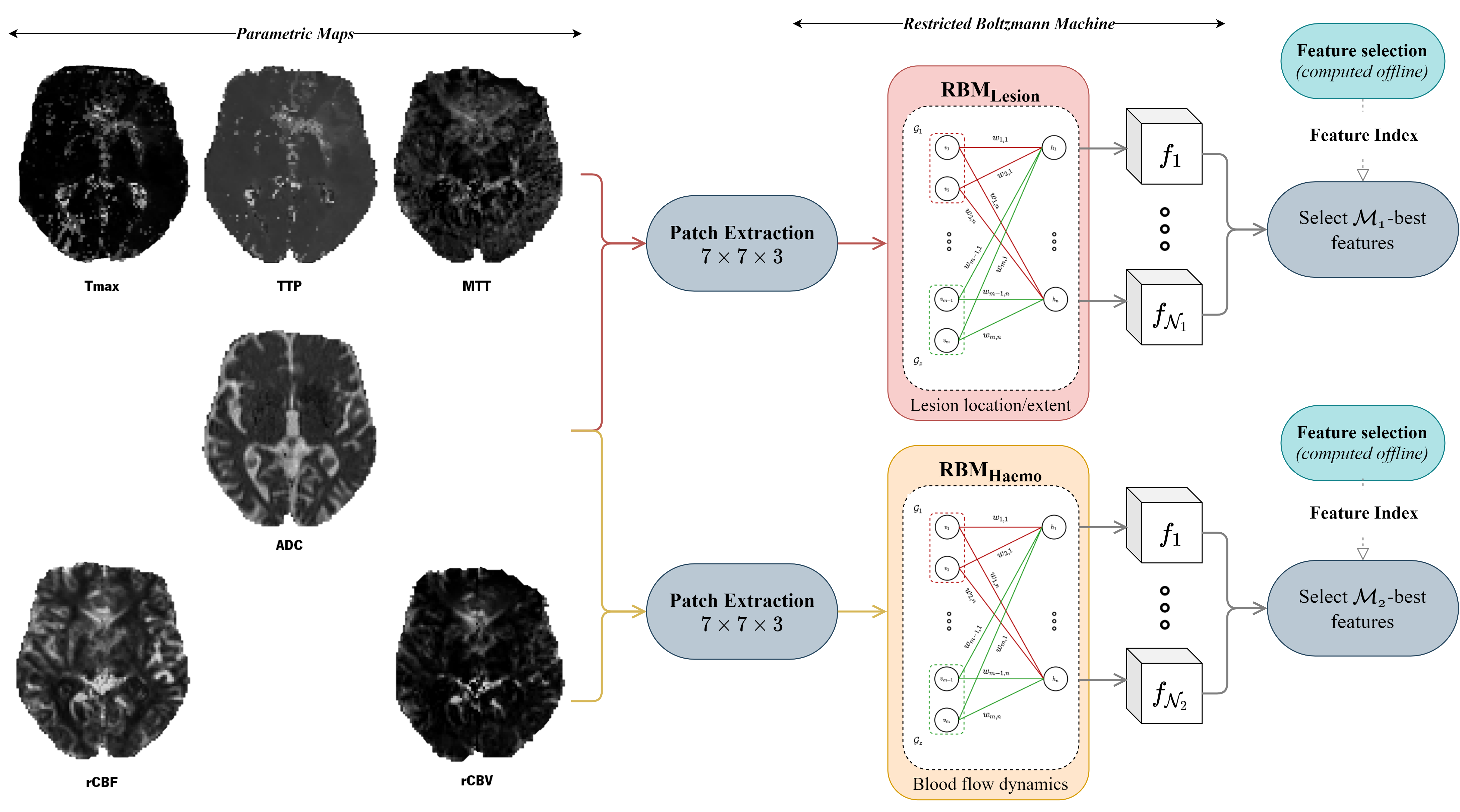}
			\centering
			\caption{Overview of the proposed unsupervised learning block. For each RBM of the unsupervised learning block, the selected features were $\mathcal{M}_1 = \mathcal{M}_2 = 6$.}
			\label{fig:RBM_pipeline}
	\end{figure*}
	
	\subsection{Restricted Boltzmann Machines}
	\label{ssec:RBM_method}
	
		The RBM is an undirected graphical model constituted by two layers of nodes: a visible layer and a hidden layer \citep{rumelhart1986parallel}. Each node has a weighted connection to all nodes in the other layer \citep{rumelhart1986parallel}. However, there are no connections among nodes of the same layer. Originally, \citet{rumelhart1986parallel} proposed the RBM to learn from binary data on both layers. However, this does not represent well continuous real-valued input data, which is the case of MRI data. Therefore, we model the visible nodes as linear units with independent Gaussian noise. The hidden nodes are modelled as Noisy Rectifier Linear Units (NReLU), since they have been reported to be suitable for feature extraction \citep{hinton2012practical}. This kind of RBM was previously used in segmentation tasks, such as in \citet{pereira2018adaptive}. Mapping the input data into a feature vector is performed through the interaction of states between the visible and hidden units, which is learned by minimizing an energy function.
			
		The complete pipeline of the unsupervised block is shown in Figure \ref{fig:RBM_pipeline} and detailed in Section \ref{ssec:Model_config}. The RBM\textsubscript{Lesion} and RBM\textsubscript{Haemo} function as feature generators that output two complementary sets of feature maps $\mathcal{N}_1$ and $\mathcal{N}_2$. These features characterise the structure of the images; however, we are interested only on the most distinctive details. So, after training the RBMs, we perform feature selection to reduce the generated feature space, obtaining smaller but representative feature sets $\mathcal{M}_1$ and $\mathcal{M}_2$, such that $| \mathcal{M}_i| \ll |\mathcal{N}_i|$, for $i \in [1, 2]$, where the operator $|.|$ denotes the cardinality of a set. In the feature selection, we would like to select the features from the RBM that encodes the MRI maps, but also that correlates with the stroke prediction. Since the RBM is an unsupervised method, we compute the Normalized Mutual Information to quantify the statistical dependence between each generated feature and the respective input MRI map, as defined by Equation \ref{eq:nmi_eq} \citep{vinh2010information}:

		\begin{equation}
			NMI_{sum} (MRI_x, Feat_y) = 2 \frac{MI(MRI_x, Feat_y)}{H(MRI_x) + H(Feat_y)},
			\label{eq:nmi_eq}
		\end{equation}
		
		\noindent
		where $MI(.)$ is the mutual information between an MRI parametric map, $MRI_x$, and an output feature, $Feat_y$; $H(.)$ defines the entropy of a map, namely, $MRI_x$ and $Feat_y$. To relate the features of the RBM with the class label, we could use a classifier supervisedly trained. Since the neural network is trained iteratively, we use a RF classifer trained with the Mean Decrease Impurity (MDI) as a surrogate to make the feature selection tractable. After, we compute the MI\textsubscript{RBM} and and MDI\textsubscript{RF}, normalize the MI\textsubscript{RBM} by the maximum value, add both ranks and sort decreasingly. The best set will be the first $\mathcal{M}_i$ features. Our selection method was inspired on the work of \citet{pereira2018enhancing}; however, their method cannot be directly applied, since it would generate too many features for our problem.
		
	\subsection{Convolutional and Recurrent Neural Networks}
			
		Our supervised functional block is based on the U-Net architecture as proposed by \citet{ronneberger2015u}. The input of the U-Net considers the concatenation of standard parametric maps with the sets of feature maps extracted from the unsupervised block. In the first level of our encoder architecture we use four 2D convolutional blocks with kernel size of $3 \times 3$ and $32$ channels. Afterwards, the output of the final convolutional block is down-sampled by a factor of 2, starting the second encoding level formed by two convolutional blocks with equal kernel size but doubling the number of feature maps. The third level of encoding follows the same pattern. The decoder level mimics the encoder counterpart. As in \citet{ronneberger2015u} we only used long skip connections among encoder and decoder levels. These encoder-decoder deep CNNs provide high levels of abstraction from the input data, increasing the global notion of context as the network grows deeper. However, it comes at a cost of a high receptive field \citep{zeiler2014visualizing}. Thus, we used a 2D architecture in the plane with the highest resolution, since the acquisition resolution is anisotropic in the dataset. Also, in the end of the decoding path we expanded our learning block with Gated RNNs. Due to their nature, Gated RNNs can capture short- and long-term spatial relations, by retaining information from previous nodes encoded in the time-steps. Hence, Gated-RNNs consider information from all previous nodes when analysing the current one. This property, when applied to imaging data, allows considering intra-slice contextual dependencies. In our work, we used a particular Gated-RNN, namely the Long-Short Term Memory (LSTM) \citep{hochreiter1997long}. However, the LSTM was intrinsically developed to process 1D data \citep{hochreiter1997long} (\eg time-series). To be applicable to 2D data, we developed an online 2D Partition layer that transforms a grid-structure input (\eg an image) into a one-dimensional sequence. Inspired by \citet{visin2016reseg}, the 2D Partition layer was predefined with a neighbourhood of $2 \times 2$, where each time-step is characterised by a feature space of four voxels. After, two Bidirectional LSTM layers are employed along the left-right and frontal-dorsal directions followed by an up-sampling layer. These four layers, referred as the Gated Recurrent block, are shown in Figure \ref{fig:architecture}. In our supervised functional block, two Gated Recurrent blocks were used, where the Bidirectional LSTMs have $64$ and $32$ hidden layers, respectively. The impact of the main components is evaluated in an ablation study in the experiments.
			
\section{Experimental Setup}
\label{sec:eval}
	
	We evaluated the proposed approach on the publicly available ISLES 2017 dataset and on a private dataset. ISLES has an online benchmark platform \citep{kistler2013virtual} that performs automatic evaluation \citep{Site}. In this section we describe the dataset, the training and evaluation, and the main hyper-parameters of our method.

	\subsection{Data} 

		ISLES 2017 dataset encompasses 75 ischaemic stroke patients, which are separated into two sets: training ($n=43$) and testing ($n=32$). Both sets have patients that underwent mechanical thrombectomy. Each patient is characterised by six 3D parametric MRI maps: diffusion ADC map, perfusion rCBF, rCBV, TTP, MTT and Tmax maps. In addition to the standard parametric maps, each case is also characterised by a manual delineation of the lesion. This refers to the 90-day stroke lesion delineated with access to the follow-up T2-weighted acquisition. However, the manual delineation is only available for the training set, while the follow-up T2-weighted imaging is not disclosed for any set.  All parametric MRI maps are already co-registered and skull-stripped \citep{winzeck2018isles}. Figure \ref{fig:rbm_maps} (top row) shows an example of MRI maps, alongside the manual lesion delineation, the Ground Truth (GT), of a patient.
		
		The private dataset considers 23 acute ischaemic stroke patients that underwent clinical therapy, acquired at Bern University Hospital in Switzerland. As in ISLES 2017, each patient is characterized by the same six parametric maps, being the final lesion manually delineated at 90-day follow-up T2. The parametric maps were co-registered followed by skull-stripping with FSL BET2 on the co-registered follow-up T2 image \citep{jenkinson2012fsl}.

	\subsection{Evaluation Metrics}

		We evaluated our proposal with five metrics, which are the same ones computed by the online ISLES 2017 benchmark platform: Dice Similarity Score, Hausdorff Distance (HD), Average Symmetric Surface Distance (ASSD), Precision, and Recall \citep{kistler2013virtual}.
	
		Dice score measures the spatial overlap between two volumes. HD corresponds to the highest distance between surface points of different volumes, which characterise spatial outliers in the prediction. ASSD quantifies the average distances between the volumes' surface. Precision quantifies the proportions of correctly classified cases within a class, while Recall corresponds to the proportion of positive cases correctly identified as such.
		
	\subsection{Image pre- and post-processing}
		
		Since MRI acquisitions were acquired from different centers and configurations \citep{winzeck2018isles}, for each patient we resized all maps to a common volume of dimension of $256 \times 256 \times 32 $. Afterwards, the ADC maps were clipped between $\left[ 0, 2600 \right] \times 10^{-6} mm^2/s$ and the Tmax maps were clipped to $ \left[ 0, 20s \right]$, since values beyond these ranges are known to be biologically meaningless \citep{mckinley}. Finally, a linear scaling was applied across all maps, to the range $\left[ 0, 255 \right]$. The images are resized to its original size, after we perform the prediction.

		We applied a morphological filtering as post-processing, but since the final stroke lesion presents a wide range of lesion volumes \citep{winzeck2018isles}, we removed only small connected components with less than 250 voxels. This step was kept fixed for all the evaluated models.

	\subsubsection{Data Augmentation}

		Data augmentation can be used to increase the number of training samples and reduce over-fitting \citep{krizhevsky2012imagenet}. Due to the relatively small size of the training dataset, we employed artificial data augmentation in the supervised portion of our proposal. For each sample, we applied rotations of $90\degree$, $180\degree$, $270\degree$.
		
		\subsection{Settings and model training}
		\label{ssec:Model_config}
		
		\paragraph{Unsupervised functional block}
		The unsupervised functional block was trained by optimizing the negative log-likelihood of the data. However, since computing the gradient is generally intractable, we performed the training by approximating the gradient with Contrastive Divergence with one step of alternating Gibbs sampling \citep{hinton2012practical}. The training process of an RBM can be difficult if one tries to learn the parameter $\sigma_i$ of the energy function, which corresponds to the standard deviation of the Gaussian noise of a visible node $i$ \citep{hinton2012practical}. According to \citet{hinton2012practical}, we normalize each component of the data with zero mean and unit variance, and define $\sigma_i = 1$. In Table \ref{tab:params}, we present the settings used for training the unsupervised model. 
		
		For training each RBM, we randomly extract 3D patches of shape $ 7 \times 7 \times 3$ from the respective input set of MRI maps, $\mathcal{C}$. Then, the 3D patches are reshaped into a 1D vector and fed into the visible layer of the RBM, having an input of size $m= 7 \times 7 \times 3 \times |\mathcal{C}|$, as shown in Figure \ref{fig:RBM_pipeline}. After training, we extract features from the NReLU units noise-free activations. These units exhibit intensity equivariance when the bias has zero value, and they are noise free units \citep{nair2010rectified}. Due to the large number of extracted feature maps ($ |\mathcal{N}_1| = |\mathcal{N}_2| = 600 $), we perform a feature selection step, as described in Section \ref{ssec:RBM_method}, where $\mathcal{M}_1 = \mathcal{M}_2 = 6$. The most appropriate cardinality of $\mathcal{M}$ is discussed in Section \ref{sssec:unsupervised}.
		
		\begin{table}[H]
	\renewcommand\thetable{1}
    \caption{Model training parameters for the unsupervised and supervised functional blocks.}
    \flushleft
    \centering
    \resizebox{0.45\textwidth}{!}{
        \begin{tabular}{ccc}
            \specialrule{.2em}{.1em}{.1em}
            Functional Block & Parameter & Description              \\
            \cmidrule(lr){1-3}
            
            \multirow{3}{*}{Unsupervised} 
            & Optimizer & SGD with momentum ($lr=1\times10^{-5}$)  \\
            & Patch shape   & $7 \times 7 \times 3$      \\
            & Batch size   & $32$ \\
            
            \addlinespace[0.06cm]
            \cmidrule(lr){1-3}
            \addlinespace[0.06cm]

            \multirow{3}{*}{Supervised} 
            & Optimizer  & ADAM ($lr=1\times10^{-5}$) \\
            & Patch shape & $84 \times 84$ \\
            & Batch size & $4$ \\
            \bottomrule                    
        \end{tabular}
    }
    \label{tab:params}
\end{table}
		
		\paragraph{Supervised functional block}
		As for the supervised functional block, the complete settings of the training are given in Table \ref{tab:params}. For each subject, $350$ patches were randomly sampled. The training comprehended $36$ subjects, while the remaining $7$ subjects were used for validation. The settings were optimized through cross-validation in a previous work \citep{pinto2018stroke}. For training, we used soft Dice loss function \citep{milletari2016v}. It is defined as:

		\begin{equation}
			\text{Soft Dice loss} = \frac{\sum_{i}^{|V|} p_i g_i}{\sum_{i}^{|V|} p_{i}^{2} + \sum_{i}^{|V|} g_{i}^{2}}.
			\label{eq:f_beta}
		\end{equation}
		
		In the soft dice loss, the sum occurs over the set $V$ of voxels belonging to the predicted output patch, where $p_i \in \mathcal{P}$ denotes the probability of a voxel $i$ in the output patch and $g_i \in \mathcal{G}$ corresponds to the respective ground-truth label voxel.
			
		The method was implemented using Keras with Tensorflow backend, in a workstation equipped with a GTX 1080 Ti 11 GB. Prediction time takes around 20s per patient.\footnote{Additional details of setting and model training are provided in the supplementary material. Also, the source code for reproducing the segmentations, the models’ weights and segmentations can be found at: \url{https://github.com/apinto92/stroke_prediction.git}.}
		

\section{Results and Discussion}
\label{sec:results}
	
	In this section, we discuss the impact of the main contributions, namely, the incorporation of unsupervised learning with supervised learning and the Gated Recurrent blocks. Then, we compare our method with the state of the art in ISLES 2017 Challenge. Finally, we delve on the difficulty of predicting the final infarct stroke lesion.
	
	\subsection{Ablation Study}
	\label{sssec:unsupervised}
		The ablation study aims to gradually measure the importance of the main components and consequently assert on the contribution of each component to the overall performance. Thus, we start by evaluating the importance of the unsupervised feature generator and the proposed input grouping. After, we focus on the use of the Gated Recurrent block and the choice of the dimensionality of the spatial context.
		
		\makeatletter
\g@addto@macro{\endtabular}{\rowfont{}}
\makeatother
\newcommand{\rowfonttype}{}
\newcommand{\rowfont}[1]{
\gdef\rowfonttype{#1}#1\ignorespaces%
}
\makeatother

\begin{table*}[!hb]
	\renewcommand\thetable{2}
    \caption{Results obtained with different configurations of the unsupervised feature generator block in ISLES 2017 testing set. Each metric represents the mean $\pm$ standard deviation. Underlined values correspond to the highest mean.}
    \tiny
    \centering
    \resizebox{\textwidth}{!}{

    \begin{threeparttable}
    \centering
    \begin{tabular}{c|cc|c|c|c|c|c}
            \specialrule{.2em}{.1em}{.1em}
            \multirow{2}{*}{Unsupervised Block}
            & \multicolumn{2}{c|}{Supervised Block}
            & \multirow{2.5}{*}{Dice}
            & \multirow{2.5}{*}{HD}
            & \multirow{2.5}{*}{ASSD}
            & \multirow{2.5}{*}{Precision}
            & \multirow{2.5}{*}{Recall}
            \\[0.08cm]
            
            & FCN & G-RNN & & & & & \\
            \cline{1-8}
            
            \multicolumn{1}{c|}{\vline height10pt width0pt\relax -- } & U-Net & LSTM & 0.30 $\pm$ 0.21 & 36.58 $\pm$ 16.62 & 6.96 $\pm$ 5.08 & 0.30 $\pm$ 0.26 & 0.55 $\pm$ 0.31 \\[0.1cm]

            \cline{1-8}                                 
            
            \multicolumn{1}{c|}{\vline height10pt width0pt\relax RBM\textsubscript{Single} (3 Feat.)} & U-Net & LSTM & 0.30 $\pm$ 0.21 & 38.93 $\pm$ 18.80 & 6.55 $\pm$ 4.22 & 0.29 $\pm$ 0.24 & 0.61 $\pm$ 0.31 \\[0.1cm]
            
            RBM\textsubscript{Single} (6 Feat.) & U-Net & LSTM & 0.30 $\pm$ 0.21 & 36.94 $\pm$ 19.19 & 6.72 $\pm$ 4.43 & 0.29 $\pm$ 0.24 & 0.59 $\pm$ 0.31 \\[0.1cm]
            
            RBM\textsubscript{Single} (12 Feat.) & U-Net & LSTM & 0.28 $\pm$ 0.20 &  41.07 $\pm$ 18.67 & 6.81 $\pm$ 3.88 & 0.24 $\pm$ 0.21 & 0.65 $\pm$ 0.30 \\[0.1cm]
            
            \cline{1-8}
             
            \multicolumn{1}{c|}{\vline height10pt width0pt\relax RBM\textsubscript{Haemo}} & U-Net & LSTM & 0.28 $\pm$ 0.24 & 38.50 $\pm$ 22.78  & 11.09 $\pm$ 14.79 & 0.35 $\pm$ 0.30 & 0.44 $\pm$ 0.34  \\[0.1cm]
            
            RBM\textsubscript{Lesion} & U-Net & LSTM & 0.31 $\pm$ 0.21 & 35.38 $\pm$ 15.75 & 6.44 $\pm$ 4.43 & 0.30 $\pm$ 0.24 & 0.59 $\pm$ 0.30 \\[0.1cm]
            
             \cline{1-8}
             
            \multicolumn{1}{c|}{\vline height10pt width0pt\relax RBM\textsubscript{Lesion} $+$ RBM\textsubscript{Haemo}} &  U-Net & LSTM & \underline{0.38} $\pm$ 0.22 & \underline{29.21} $\pm$ 15.04 & \underline{5.52} $\pm$ 5.06 & \underline{0.41} $\pm$ 0.26 & 0.53 $\pm$ 0.29 \\[0.1cm]

            Two-RBMs\textsubscript{Mixed} & U-Net & LSTM & 0.27 $\pm$ 0.21 & 40.89 $\pm$ 14.63 & 6.92 $\pm$ 3.64 & 0.25 $\pm$ 0.23 & \underline{0.68} $\pm$ 0.28 \\[0.1cm]

            Three-RBMs\tnote{1} &  U-Net &  LSTM &  0.35 $\pm$ 0.23 & 29.32 $\pm$ 14.33 &  5.27 $\pm$ 3.54 &  0.34 $\pm$ 0.27 &  0.59 $\pm$ 0.30 \\
            
            \bottomrule
    \end{tabular}
    	\begin{tablenotes}
         	\item[1] RBM\textsubscript{Haemo/Less} + RBM\textsubscript{Lesion/Less} + RBM\textsubscript{ADC}
       \end{tablenotes}
    \end{threeparttable}}

\label{tab:RBM_ablation}
\end{table*}

		\subsubsection{Unsupervised feature generation}
		\label{sssec:unsupervised}
			
\begin{figure*}[!ht]
				\includegraphics[width=\textwidth]{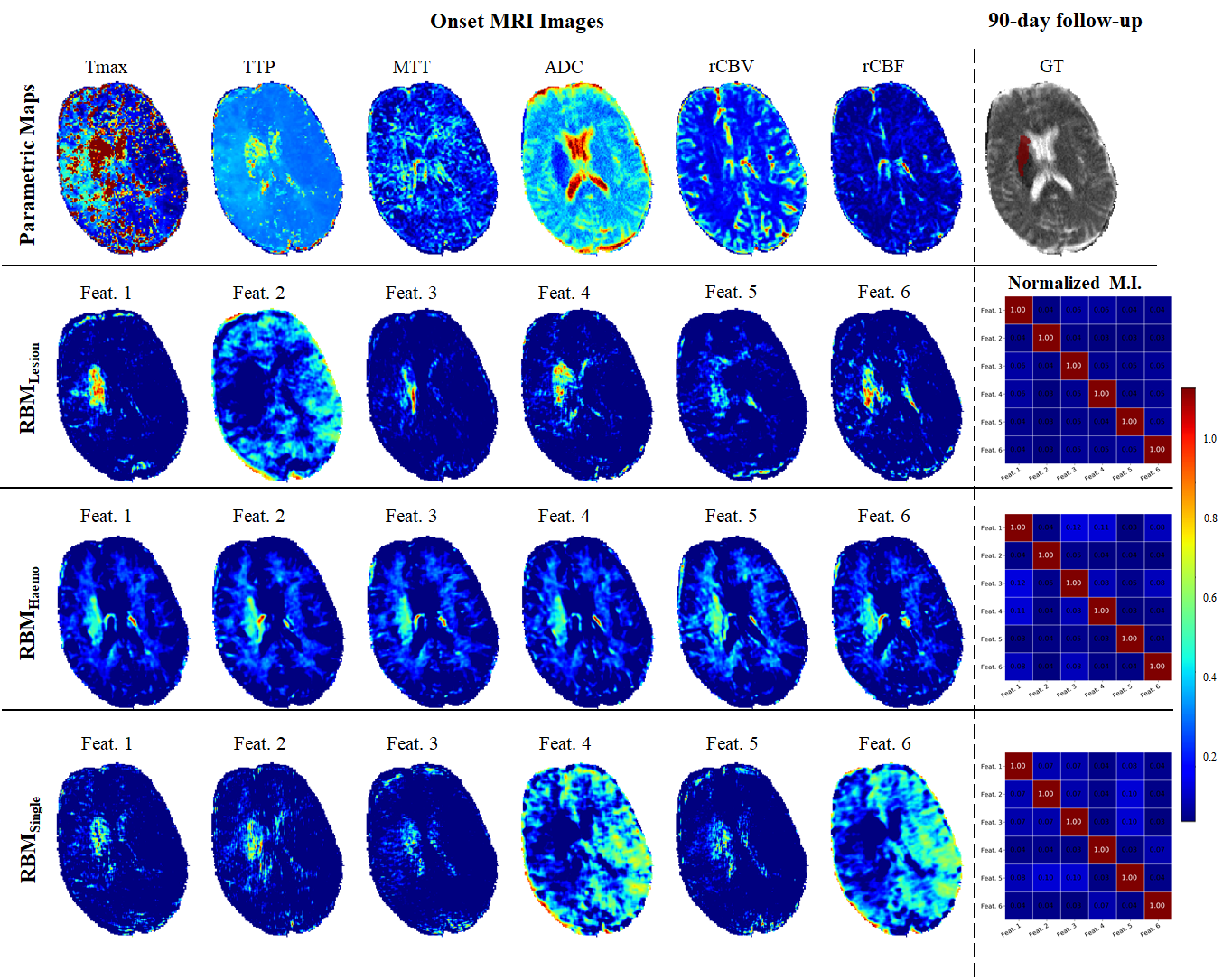}
				\centering
				\caption{Onset parametric maps of patient case $0011$ in ISLES 2017 training set, alongside the final stroke lesion, at a 90-day follow-up, over the onset ADC map. The subsequent rows show the RBM features selected from the RBM\textsubscript{Lesion}, RBM\textsubscript{Haemo} and RBM\textsubscript{Single}, respectively. The last column shows the normalized mutual information, across whole dataset, among features of the same RBM.}
				\label{fig:rbm_maps}
			\end{figure*}

			We hypothesize that grouping the parametric MRI maps according to their physical meaning and encoding each group with an RBM has potential to extract better features to characterise the stroke lesion and the blood haemodynamics. We perform several experiments to corroborate this working hypothesis. In all experiments, the parametric MRI maps are also used as input to the supervised block. The results are presented in Table \ref{tab:RBM_ablation}. Figure \ref{fig:rbm_maps} presents feature maps encoded by the RBMs and the respective MRI maps.

			\paragraph{Grouping all parametric MRI maps in a single group}
		
			We considered, first, the effect of encoding all parametric maps using a single RBM. We varied the number of selected features from the RBM, observing that in all cases, the average Dice score is equal or lower than using only the parametric maps as input to the supervised block. Also, using $12$ features presented the lowest average Dice score. The use of $3$ or $6$ obtained the same average Dice score, having the second a lower average Hausdorff distance. So, based on the metrics, we may conclude that there is no clear gain in using the features generated by the RBM, at least, when we encode all the parametric maps with a single RBM.

			Since, the selection of $6$ features also includes the previous top $3$ features, we compared the normalized mutual information between them. As shown in Figure \ref{fig:rbm_maps}, the top $3$ features have low values of normalized mutual information in relation to the additional $3$ features, which indicates that there is additional information. For this reason, we chose $6$ as the number of features in the subsequent experiments.

			\paragraph{Grouping parametric MRI maps according to the subjacent physical meaning}
			
				In this experiment, we grouped the parametric maps according to their underlining physical meaning together with ADC map in each group. Each group was encoded with an RBM. Comparing isolatedly the use of each group of features, we verify that RBM\textsubscript{Lesion} had a higher average Dice score compared to using only the parametric maps as input to the supervised block. The increase in the average Dice score was obtained by a higher average Recall. Also, we observe an improvement in all distance metrics. The experiment of using RBM\textsubscript{Haemo} presented the lowest average Dice and Recall, as well as higher average distance metrics. However, RBM\textsubscript{Haemo} presented higher average Precision, contrary to RBM\textsubscript{Lesion}, which motivated the study on the combination of features from RBM\textsubscript{Haemo} with RBM\textsubscript{Haemo} besides the parametric maps. The results of this experiment are presented in Table \ref{tab:RBM_ablation}. We may observe that this combination obtained the highest average Dice and Precision, as well as the lowest average distance metrics. However, this improvement could have been originated from the combination of maps according to a specific common property, subjacent physical meaning of the parametric maps, in each group, or because we reduced the number of maps from $6$ to $3$ in each group. And this reduction could have allowed a better training of the RBM. So, we performed two complementary experiments. In the first experiment, we formed two groups with similar size, but we randomly chose the parametric maps to include in each group. In the second experiment (Three-RBMs), we changed the groups of MRI maps encoded in RBM\textsubscript{Lesion} and RBM\textsubscript{Haemo} by removing the ADC map from each one. These two new groups were encoded in RBM\textsubscript{Lesion/Less} and RBM\textsubscript{Haemo/Less}, respectively. The ADC was separately encoded in RBM\textsubscript{ADC}. As presented in Table \ref{tab:RBM_ablation}, the first experiment presented the lowest average Dice score and higher average distance metrics, while the second experiment attained the second highest average Dice score, thus showing the importance of splitting the parametric perfusion maps and including the ADC map in both the RBM\textsubscript{Haemo} and RBM\textsubscript{Lesion}.
				
			\begin{table*}[!hb]
    \tiny
    \caption{Results obtained when considering the Gated Recurrent block with and without the unsupervised learning block with ISLES 2017 testing set. Each metric represents the mean $\pm$ standard deviation. Underlined values correspond to the highest mean.}
    \centering
    \resizebox{\textwidth}{!}{
    \begin{tabular}{c|cc|c|c|c|c|c}
        \specialrule{.2em}{.1em}{.1em}
        \multirow{2.5}{*}{Unsupervised Block}
        &\multicolumn{2}{c|}{Supervised Block}
        & \multirow{2.5}{*}{Dice}
        & \multirow{2.5}{*}{HD}
        & \multirow{2.5}{*}{ASSD}
        & \multirow{2.5}{*}{Precision}
        & \multirow{2.5}{*}{Recall}
        \\[0.08cm]
        
        & FCN & G-RNN & & & & & \\
        \cline{1-8}
        
         \multicolumn{1}{c|}{\vline height10pt width0pt\relax \multirow{2}{*}{--} } & U-Net\phantom{\dag} & -- & 0.30 $\pm$ 0.21 & 38.83 $\pm$ 21.10 & 7.08 $\pm$ 5.15 & 0.26 $\pm$ 0.23 & \underline{0.64} $\pm$ 0.30 \\[0.1cm]

         & U-Net\phantom{\dag} & LSTM & 0.30 $\pm$ 0.21 & 36.58 $\pm$ 16.62 & 6.96 $\pm$ 5.08 & 0.30 $\pm$ 0.26 & 0.55 $\pm$ 0.31 \\[0.1cm]

        \cline{1-8}
        
        \multicolumn{1}{c|}{\vline height10pt width0pt\relax \multirow{2}{*}{RBM\textsubscript{Lesion} $+$ RBM\textsubscript{Haemo} [3D]}} & U-Net\phantom{\dag} & -- & 0.32 $\pm$ 0.23 & 34.09 $\pm$ 16.51 & 7.60 $\pm$ 7.14 & 0.35  $\pm$ 0.27 & 0.48 $\pm$ 0.32 \\[0.1cm]

        & U-Net\phantom{\dag} & LSTM & \underline{0.38} $\pm$ 0.22 & \underline{29.21} $\pm$ 15.04 & \underline{5.52} $\pm$ 5.06 & \underline{0.41} $\pm$ 0.26 & 0.53 $\pm$ 0.29 \\
         
        \bottomrule                         
    \end{tabular}}
    
    \label{tab:Training_ablation}
\end{table*}
			
				Considering these experiments together, we may draw some conclusions. First, although CNNs are very effective in generating features from raw data, they can generate even better features if rich and complementary information is provided. A similar conclusion was inferred by \citet{oliveira2018retinal} that observed improvement when the coefficients of the Wavelet were added as input in the problem of retinal vessel segmentation. Here, we observe a similar effect, but using the encoding provided by an RBM trained unsupervisedly for the problem of stroke lesion prediction. Second, at least to the problem of stroke lesion prediction, when we have data with different latent factors and we are able to group it, according to those factors, then there is potential to extract complementary information from each group, but to mix them all together can be detrimental.	
			
			\subsubsection{Context aggregation based on gated recurrent blocks}

			In medical imaging segmentation, which is similar to our problem of inferring the extension of the lesion 90 days ahead, the use of a cascade of convolutional layers to elaborate the features is the prevalent practise. However, as discussed previously, Gated-RNN layers are able to capture long distance spatial relations among input voxels, so we performed some experiments to evaluate its contribution. The results are presented in Table \ref{tab:Training_ablation}.
			
			\begin{table*}[!ht]
    \tiny
    \caption{Evaluation metrics obtained with different spatial context configurations in the unsupervised and supervised learning blocks in ISLES 2017 testing set. Each metric represents the mean $\pm$ standard deviation. Underlined values correspond to the highest mean.}
    \centering
    \resizebox{\textwidth}{!}{
    \begin{tabular}{c|cc|c|c|c|c|c}
        \specialrule{.2em}{.1em}{.1em}
        \multirow{2.5}{*}{Unsupervised Block}
        &\multicolumn{2}{c|}{Supervised Block}
        & \multirow{2.5}{*}{Dice}
        & \multirow{2.5}{*}{HD}
        & \multirow{2.5}{*}{ASSD}
        & \multirow{2.5}{*}{Precision}
        & \multirow{2.5}{*}{Recall}
        \\[0.08cm]
        
        & FCN & G-RNN & & & & & \\
        \cline{1-8}

        \multicolumn{1}{c|}{\vline height10pt width0pt\relax RBM\textsubscript{Lesion} $+$ RBM\textsubscript{Haemo} [2D]} & U-Net [2D] & -- & 0.27 $\pm$ 0.23 & 36.35 $\pm$ 14.89 & 9.14 $\pm$ 12.35 & 0.31  $\pm$ 0.28 & 0.53 $\pm$ 0.34 \\[0.1cm]

        \cline{1-8}
        
        \multicolumn{1}{c|}{\vline height10pt width0pt\relax \multirow{2}{*}{RBM\textsubscript{Lesion} $+$ RBM\textsubscript{Haemo} [3D]}} & U-Net [2D] & -- & 0.32 $\pm$ 0.23 & 34.09 $\pm$ 16.51 & 7.60 $\pm$ 7.14 & 0.35  $\pm$ 0.27 & 0.48 $\pm$ 0.32 \\[0.1cm]
        
        & U-Net [3D] & -- & 0.30 $\pm$ 0.21 & 34.17 $\pm$ 14.86 & 6.16 $\pm$ 3.82 & 0.32 $\pm$ 0.27 & 0.54 $\pm$ 0.30 \\[0.1cm]
         
        \bottomrule                          
    \end{tabular}}
    \label{tab:Training_2D3D}
\end{table*}	
			
\begin{table*}[!hb]
    \tiny
    \caption{Results obtained by our proposal and baseline method in the private dataset. Each metric is represented by the mean $\pm$ standard deviation. Underlined values correspond to the highest mean, while bold values represent statistically significant values (\textit{p-value} $< 0.05$).}
    \centering
    \resizebox{\textwidth}{!}{
    \begin{tabular}{c|cc|c|c|c|c|c}
        \specialrule{.2em}{.1em}{.1em}
        \multirow{2.5}{*}{Unsupervised Block}
        &\multicolumn{2}{c|}{Supervised Block}
        & \multirow{2.5}{*}{Dice}
        & \multirow{2.5}{*}{HD}
        & \multirow{2.5}{*}{ASSD}
        & \multirow{2.5}{*}{Precision}
        & \multirow{2.5}{*}{Recall}
        \\[0.08cm]
        
        & FCN & G-RNN & & & & & \\
        \cline{1-8}

        \multicolumn{1}{c|}{\vline height10pt width0pt\relax --} &  U-Net [2D] &  -- &  0.32 $\pm$ 0.18 &  32.58 $\pm$ 20.09 &  5.17 $\pm$ 3.34 &  0.31 $\pm$ 0.26 &  0.68 $\pm$ 0.28 \\[0.1cm]

        \cline{1-8}
        
        \multicolumn{1}{c|}{\vline height10pt width0pt\relax RBM\textsubscript{Lesion} $+$ RBM\textsubscript{Haemo} [3D]} &  U-Net [2D] &  LSTM &  \textbf{\underline{0.36}} $\pm$ 0.18 &  \underline{26.68} $\pm$ 15.60 &  \underline{3.88} $\pm$ 2.17 &  \textbf{\underline{0.38}} $\pm$ 0.30 &  0.68 $\pm$ 0.27 \\[0.1cm]

        \bottomrule
                                        
    \end{tabular}}
    \label{tab:BERN2020}
\end{table*}
			
			Analysing Table \ref{tab:Training_ablation}, we verify that when we just had parametric maps as input to the supervised block, adding a LSTM layer increased the average Precision, but the average Recall decreased, resulting in the same average Dice score. But, a different behaviour is observed when we added the features computed from the RBMs. In this scenario, we verify that using only CNN layers improved over having just parametric maps, which came by a higher average Precision. However, when we add the LSTM, we have an even higher improvement, which is observed in a larger increase in the average Precision, and a decrease in the average distance metrics.
			
			Based on these experiments, we may conclude that the CNN layers were able to extract additional information from the RBM features; however, at least to the problem of inferring the extension of the lesion months ahead, long and local distance spatial relations among input voxels introduced by Gated RNN was critical to reduce the detection of false positives, increasing substantially the average Dice score by $6\%$.
			
			\begin{table*}[!ht]
	\tiny
	\caption{Published methods in ISLES 2017 testing dataset and our proposal. Each metric is represented by the mean $\pm$ standard deviation. Underlined values correspond to the highest mean.}
	\centering
	\resizebox{\textwidth}{!}{

	\begin{tabular}{ccccccc}
		\specialrule{.2em}{.1em}{.1em}
		&
		& \multicolumn{1}{c}{Dice} 
		& \multicolumn{1}{c}{HD} 
		& \multicolumn{1}{c}{ASSD}
		& \multicolumn{1}{c}{Precision}
		& \multicolumn{1}{c}{Recall}
		\\ \cmidrule(l){1-7}
		\multirow{4}{*}{\rotatebox[origin=c]{90}{Ensemble}}
		& Mok \etal * & 0.32 $\pm$ 0.23 & 40.74 $\pm$ 27.23 & 8.97 $\pm$ 9.52 & 0.34 $\pm$ 0.27 & 0.39 $\pm$ 0.27 \\
		& Kwon \etal * & 0.31 $\pm$ 0.23 & 45.26 $\pm$ 21.04 & 7.91 $\pm$ 7.31 & 0.36 $\pm$ 0.27 & 0.45 $\pm$ 0.30 \\
		& Robben \etal * & 0.27 $\pm$ 0.22 & 37.84 $\pm$ 17.75 & 6.72 $\pm$ 4.10 & \underline{0.44} $\pm$ 0.32 & 0.39 $\pm$ 0.31 \\
		& Pisov \etal * & 0.27 $\pm$ 0.20 & 49.24 $\pm$ 32.15 & 9.49 $\pm$ 10.56 & 0.31 $\pm$ 0.27 & 0.39 $\pm$ 029 \\
		\cmidrule(lr){1-7}
		\multirow{12}{*}{\rotatebox[origin=c]{90}{Single Model}}
		& Monteiro \etal * & 0.30 $\pm$ 0.22 & 46.60 $\pm$ 17.50 & 6.31 $\pm$ 4.05 & 0.34 $\pm$ 0.27 & 0.51 $\pm$ 0.30 \\
		& \citet{pinto2018stroke} & 0.29 $\pm$ 0.21 & 41.58 $\pm$ 22.04 & 7.69 $\pm$ 5.71 & 0.21 $\pm$ 0.21 & \underline{0.66} $\pm$ 0.29 \\
		& Lucas \etal * & 0.29 $\pm$ 0.21 & 33.85 $\pm$ 16.82 & 6.81 $\pm$ 7.18 & 0.34 $\pm$ 0.26 & 0.51 $\pm$ 0.32 \\
		& Choi \etal * & 0.28 $\pm$ 0.22 & 43.89 $\pm$ 20.70 & 8.88$\pm$ 8.19 & 0.36 $\pm$ 0.31 & 0.41 $\pm$ 0.31 \\
		
		& Niu \etal * & 0.26 $\pm$ 0.20 & 48.88 $\pm$ 11.20 & 6.26 $\pm$ 3.02 & 0.28 $\pm$ 0.25 & 0.56 $\pm$ 0.26 \\
		& Sedlar \etal * & 0.20 $\pm$ 0.19 & 58.30 $\pm$ 20.02 & 11.19 $\pm$ 9.10 & 0.23 $\pm$ 0.24 & 0.40 $\pm$ 0.29 \\
		& Rivera \etal * & 0.19 $\pm$ 0.16 & 63.58 $\pm$ 18.58 & 11.13 $\pm$ 7.89 & 0.27 $\pm$ 0.25 & 0.21 $\pm$ 0.17 \\
		& Islam \etal * & 0.19 $\pm$ 0.18 & 64.15 $\pm$ 28.51 & 14.17 $\pm$ 15.80 & 0.29 $\pm$ 0.28 & 0.25 $\pm$ 0.25 \\
		& Chengwei \etal * & 0.18 $\pm$ 0.17 & 65.95 $\pm$ 25.94 & 9.22 $\pm$ 6.99 & 0.37 $\pm$ 0.30 & 0.21 $\pm$ 0.23 \\
		& Yoon \etal * & 0.17 $\pm$ 0.16 & 45.23 $\pm$ 19.14 & 12.43 $\pm$ 11.01 & 0.23 $\pm$ 0.27 & 0.36 $\pm$ 0.32 \\
		\cmidrule(lr){2-7}
		& Baseline &  0.30 $\pm$ 0.21 & 36.58 $\pm$ 16.62 & 6.96 $\pm$ 5.08 & 0.30 $\pm$ 0.26 & 0.55 $\pm$ 0.31 \\
		
		& Proposal & \underline{0.38} $\pm$ 0.22 & \underline{29.21} $\pm$ 15.04 & \underline{5.52} $\pm$ 5.06 & 0.41 $\pm$ 0.26 & 0.53 $\pm$ 0.29 \\
		
		\bottomrule
		\noalign{\vskip 2pt}  
		\multicolumn{7}{l}{$^{*}$ Methods presented in \citet{winzeck2018isles}, whose results were retrieved from \citet{Site}.} \\
	\end{tabular}}

	\label{tab:ISLES_challenge}
\end{table*}
		
			\subsubsection{Spatial context: 2D or 3D?}

			MRI images are 3D by nature, so the use of 3D filters would allow capturing more context, which has the potential to provide better prediction. Since 2D filters are confined to a plane, unnatural discontinuous contour may occur in the perpendicular axis. However, as presented previously, the resolution of MRI images in ISLES dataset is not equal in all axis, being coarser along the axial axis. So, we studied the effect of the spatial context in our architecture. As we have two blocks, unsupervised and supervised blocks, the effect on each one was evaluated separately. The results are presented in Table \ref{tab:Training_2D3D}. Considering the results, we observe that using 2D patches in both blocks has lower average Dice score, than using only the parametric maps as input (baseline), because the increase in the average Precision was not enough to compensate the drop in the average Recall. Using 3D patches for both blocks had the same performance as our baseline. However, when we used 3D patches for the RBM but 2D blocks for the U-Net block, we improved over our baseline. This is the model with the highest average Dice score without LSTM. So, we may conclude that for our architecture, larger context using 3D patches was more effective for encoding features in the unsupervised block, while 2D patches were better suited for encoding features in the supervised U-Net-based block.
		
		\subsection{Private dataset} 

			To further evaluate the generalization capacity of our proposal, we tested it on a private dataset and compare it with the baseline method. Table \ref{tab:BERN2020} presents the results obtained by the two methods. 

			On the overall, our proposal was capable of surpassing the baseline model, attaining an higher average Dice, Precision and distance metrics, which were statistically significant (Wilcoxon Signed Ranked test with $p-value < 0.05$). Comparing to ISLES 2017 testing set, there was a slight decrease in performance. This could be explained by the shift on the intensity distribution of the MRI maps, due to different acquisition protocols or the differences in the preprocessing step.

		\subsection{State-of-the-art: ISLES 2017 Challenge} 

		The results of published methods for final infarct stroke lesion prediction using ISLES 2017 testing set \citep{winzeck2018isles}, together with our baseline and proposal methods are presented in Table \ref{tab:ISLES_challenge}. The metrics were computed by the online platform, so the ground-truth data, which was manually delineated based on a follow-up T2 MRI acquisitions, are not disclosed for public access.

		Considering the results, we observe that our baseline is competitive with an average Dice, being among the top 3 methods, and surpassing the ensemble methods of \citet{pisovneural} and \citet{robben2017dual}. Our proposed method presented the lowest distance metrics among all methods, especially for the Hausdorff distance. It obtained the second-best average Precision score, being surpassed by \citet{robben2017dual} The authors proposed the integration of meta-data information, using a two-pathway 3D network in an ensemble; however, our experiments did not indicate any improvement using 3D patches for the U-Net, at least for our architecture. So, this improvement could have come from a combination of the effect of the ensemble and the meta-data. But we note that their method presented a lower average Recall, which explains their lower average Dice score. Regarding the average Recall score, our method was fourth, but when we consider the top 3 methods, specially \citet{pinto2018stroke}, we conclude that it was obtained with a much lower average Precision, which means that to increase the true positive detections, they had to increase substantially the false positives. So, comparing with the state of the art, our method presented a better balance between Precision and Recall, which reflected into a higher average Dice score.
				
		Based on the results, we may conclude that the use of complementary features provided by the RBMs and the use of LSTM for a larger context allowed our baseline to surpass current state-of-the-art methods.
		
		\paragraph{Results from ChallengeR Benchmark}
		
		The SMIR platform of ISLES 2017 provides a weekly benchmark report of the current top-10 methods in the testing set, according to the average Dice score. So, some of the methods may not be published, lacking a description on their implementation, and, for this reason, were not included in the previous discussion. 
	
\begin{figure}[!h]
			\centering
			\includegraphics[width=0.40\textwidth]{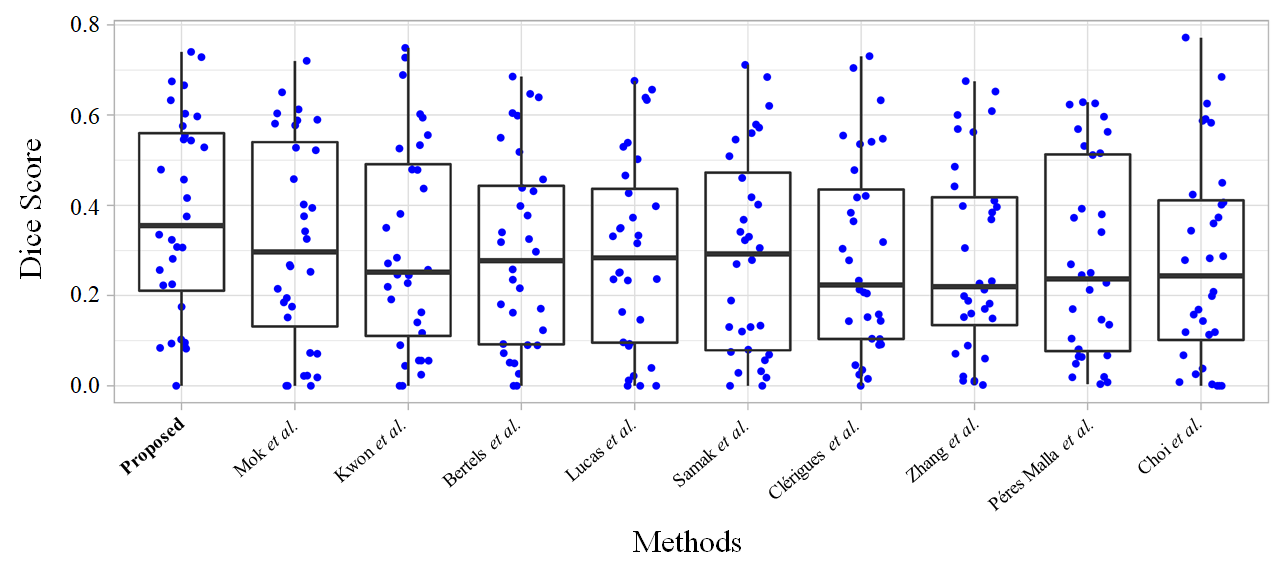}
			\caption{Boxplot of the top-10 ranking methods ordered by average Dice score in ISLES 2017 testing set.}
			\label{fig:SMIR_Boxplot}
		\end{figure}

		Figure \ref{fig:SMIR_Boxplot} presents the boxplots of each method considered in the report.
	
				\begin{figure*}[!hb]
			\renewcommand\thefigure{7}
			\centering
			\includegraphics[width=0.9\textwidth]{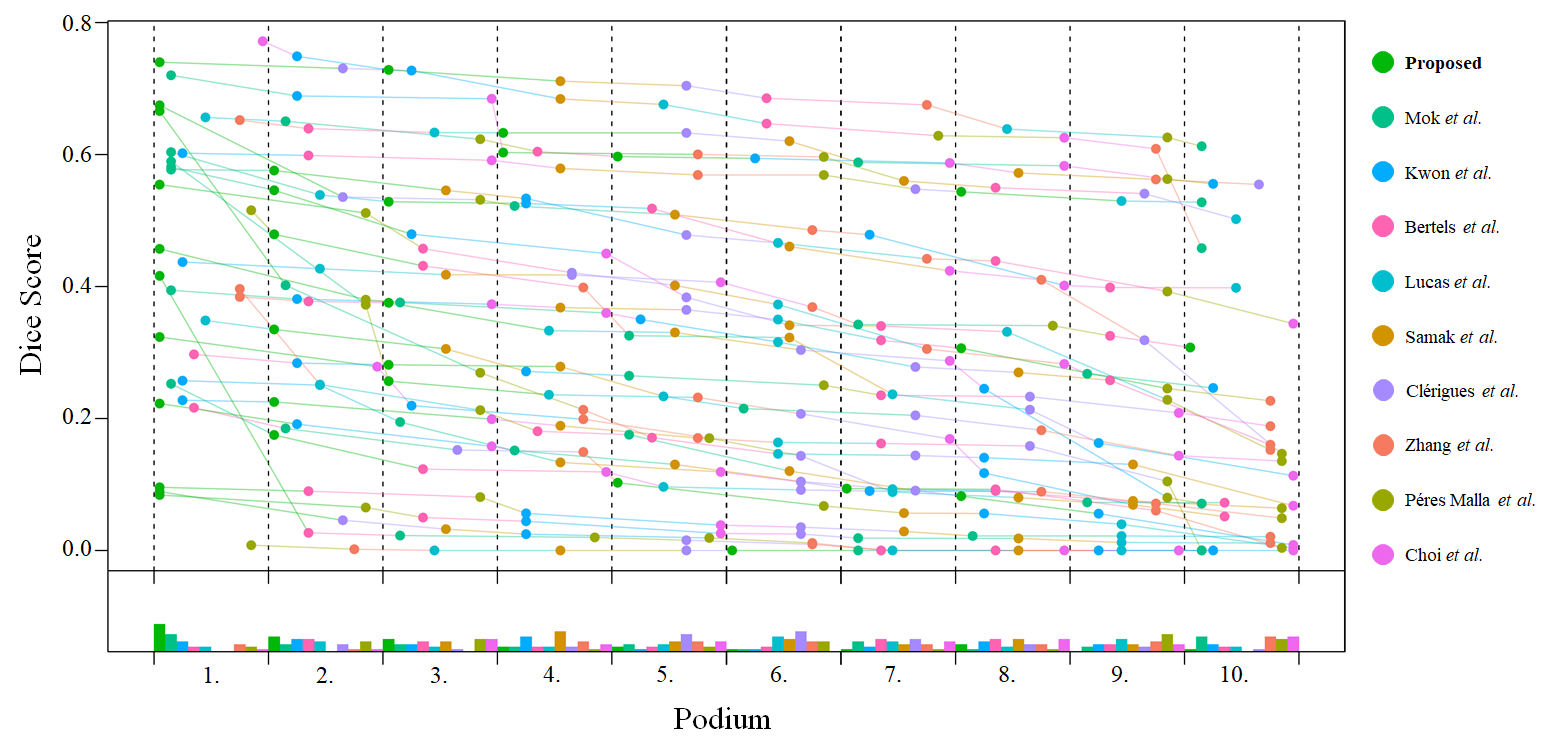}
			\caption{Podium plot of each testing case in ISLES 2017. For each ISLES 2017 testing subject, defined by a coloured line with circles, the Podium plot orders decreasingly the Dice score obtained by each of the top 10 methods that are represented by coloured circles.}
			\label{fig:SMIR_Podium}
		\end{figure*}
		
\begin{figure}[!h]
			\renewcommand\thefigure{6}
			\centering
			\includegraphics[width=0.30\textwidth]{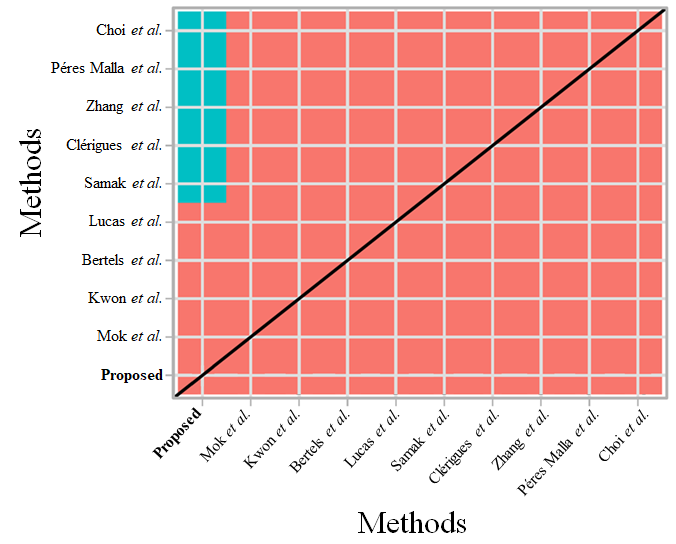}
			\caption{One-side Wilcoxon signed rank test in ISLES 2017 testing set. Statistically significant tests are marked with the blue colour, while the red colour designates statistically non-significant tests.}
			\label{fig:SMIR_Wilcox}
		\end{figure}

		We observe that the top-10 methods failed to predict the lesion of one or more cases (lowest outliers), which may indicate the degree of complexity of predicting infarct stroke lesion 90 days ahead in ISLES 2017 Challenge dataset. But we verify that our method is the only one to have the first quartile above 0.20 in the Dice score. In Figure \ref{fig:SMIR_Wilcox} we have the significance maps of the pairwise significant test with one-side Wilcoxon signed rank test (\textit{p-value} $=0.05$), showing that our method reached higher Dice score statistically significant when compared with other five ranked methods of the top-10. 
		
		Figure \ref{fig:SMIR_Podium} shows the \textit{podium} plot of each method for each case in the testing set, and its ranking. We observe that our proposal is the method, which ranked first most of the times, as well as second and third. Also, when we consider the methods ranked bellow fourth, our method is in general among those with the lowest counts. Analysing the cases individually, we note two trends, for some cases all methods presented similar performance, while for others, we find a large variation from the first to the other methods. The first trend may be found in the most difficult case, where all methods had zero or a close value for the Dice score. In the second trend, we observe that our method is ranked as first most of the cases.

		Based on the results of the benchmark, we may infer that our method is competitive among current state of the art, presenting the highest average Dice score and lowest average distance score. Considering the ablation study, this performance was attained due to the combination of adding extra features obtained by encoding the parametric maps with RBMs, according to the underlining physical meaning, and the elaboration provided by the long context of the LSTM layers.
	

\section{Conclusions}
\label{sec:conclusion}
	
	In this work, we present a deep learning approach for predicting the final stroke lesion, based on unsupervised and supervised learning. We proposed to group the input maps according to the underlying physical principle behind their creation, namely, the time-resolved perfusion maps (Tmax, TTP, MTT), and the blood-flow-dynamic related maps (rCBF, rCBV). Each group was encoded using an unsupervised model to obtain structural features specific to its underlying physical principle. These structural features together with the standard parametric maps were fed to a supervised model to learn features conditioned on the label, which in our problem, means to condition on the results of the medical intervention --- lesion at 90-days follow-up. We also investigated the use of Gated Recurrent Neural Networks to provide long spatial context, which were critical in relating the structural features to the information on input parametric maps. Our results showed that either the encoding or the long spatial context improved over our baseline. Also, these two together interacted positively increasing the performance when considering separately each one.

	When evaluating our proposal on ISLES 2017 testing dataset, we observe a prediction improvement over current state-of-the-art methods. The proposed method obtained the first place in Dice and also in HD and ASSD.
	
	Recent works \citep{pinto2018stroke,robben2020prediction} have shown the importance of clinical meta-data to predict the final stroke lesion in different revascularization scenarios. So as future work, we aim to study how such meta-data (\ie TICI score) could be incorporated in our architecture, to consolidate the impact of the clinical intervention and to further improve the 90-day lesion prediction.


\section*{Acknowledgement}

	Adriano Pinto was supported by a scholarship from the Funda\c{c}\~ao para a Ci\^encia e Tecnologia (FCT), Portugal (scholarship number PD/BD/113968/2015). This work was supported by FCT national funds, under the national support to R\&D units grant, through the reference project UIDB/04436/2020 and UIDP/04436/2020.


\bibliographystyle{model2-names}\biboptions{authoryear}
\bibliography{references}

\begin{thebibliography}{53}
\expandafter\ifx\csname natexlab\endcsname\relax\def\natexlab#1{#1}\fi
\providecommand{\url}[1]{\texttt{#1}}
\providecommand{\href}[2]{#2}
\providecommand{\path}[1]{#1}
\providecommand{\DOIprefix}{doi:}
\providecommand{\ArXivprefix}{arXiv:}
\providecommand{\URLprefix}{URL: }
\providecommand{\Pubmedprefix}{pmid:}
\providecommand{\doi}[1]{\href{http://dx.doi.org/#1}{\path{#1}}}
\providecommand{\Pubmed}[1]{\href{pmid:#1}{\path{#1}}}
\providecommand{\bibinfo}[2]{#2}
\ifx\xfnm\relax \def\xfnm[#1]{\unskip,\space#1}\fi
\bibitem[{Badrinarayanan et~al.(2015)Badrinarayanan, Kendall and
  Cipolla}]{badrinarayanan2015segnet}
\bibinfo{author}{Badrinarayanan, V.}, \bibinfo{author}{Kendall, A.},
  \bibinfo{author}{Cipolla, R.}, \bibinfo{year}{2015}.
\newblock \bibinfo{title}{Segnet: A deep convolutional encoder-decoder
  architecture for image segmentation}.
\newblock \bibinfo{journal}{arXiv preprint arXiv:1511.00561} .
\bibitem[{Bauer et~al.(2014)Bauer, Gratz, Gralla, Reyes and
  Wiest}]{bauer2014towards}
\bibinfo{author}{Bauer, S.}, \bibinfo{author}{Gratz, P.P.},
  \bibinfo{author}{Gralla, J.}, \bibinfo{author}{Reyes, M.},
  \bibinfo{author}{Wiest, R.}, \bibinfo{year}{2014}.
\newblock \bibinfo{title}{Towards automatic {MRI} volumetry for treatment
  selection in acute ischemic stroke patients}, in:
  \bibinfo{booktitle}{Engineering in Medicine and Biology Society (EMBC), 2014
  36th Annual International Conference of the IEEE},
  \bibinfo{organization}{IEEE}. pp. \bibinfo{pages}{1521--1524}.
\bibitem[{Berkhemer et~al.(2016)Berkhemer, Jansen, Beumer, Fransen, Van
  Den~Berg, Yoo, Lingsma, Sprengers, Jenniskens, Lycklama~{\`a} Nijeholt
  et~al.}]{berkhemer2016collateral}
\bibinfo{author}{Berkhemer, O.A.}, \bibinfo{author}{Jansen, I.G.},
  \bibinfo{author}{Beumer, D.}, \bibinfo{author}{Fransen, P.S.},
  \bibinfo{author}{Van Den~Berg, L.A.}, \bibinfo{author}{Yoo, A.J.},
  \bibinfo{author}{Lingsma, H.F.}, \bibinfo{author}{Sprengers, M.E.},
  \bibinfo{author}{Jenniskens, S.F.}, \bibinfo{author}{Lycklama~{\`a} Nijeholt,
  G.J.}, et~al., \bibinfo{year}{2016}.
\newblock \bibinfo{title}{Collateral status on baseline computed tomographic
  angiography and intra-arterial treatment effect in patients with proximal
  anterior circulation stroke}.
\newblock \bibinfo{journal}{Stroke} \bibinfo{volume}{47},
  \bibinfo{pages}{768--776}.
\bibitem[{Butcher and Emery(2010a)}]{butcher2010acute1}
\bibinfo{author}{Butcher, K.}, \bibinfo{author}{Emery, D.},
  \bibinfo{year}{2010}a.
\newblock \bibinfo{title}{Acute stroke imaging part i: Fundamentals}.
\newblock \bibinfo{journal}{Canadian Journal of Neurological Sciences}
  \bibinfo{volume}{37}, \bibinfo{pages}{4--16}.
\bibitem[{Butcher and Emery(2010b)}]{butcher2010acute2}
\bibinfo{author}{Butcher, K.}, \bibinfo{author}{Emery, D.},
  \bibinfo{year}{2010}b.
\newblock \bibinfo{title}{Acute stroke imaging part ii: the ischemic penumbra}.
\newblock \bibinfo{journal}{Canadian Journal of Neurological Sciences}
  \bibinfo{volume}{37}, \bibinfo{pages}{17--27}.
\bibitem[{Choi et~al.(2016)Choi, Kwon, Lee, Kim, Paik and
  Won}]{choi2016ensemble}
\bibinfo{author}{Choi, Y.}, \bibinfo{author}{Kwon, Y.}, \bibinfo{author}{Lee,
  H.}, \bibinfo{author}{Kim, B.J.}, \bibinfo{author}{Paik, M.C.},
  \bibinfo{author}{Won, J.H.}, \bibinfo{year}{2016}.
\newblock \bibinfo{title}{Ensemble of deep convolutional neural networks for
  prognosis of {Ischemic Stroke}}, in: \bibinfo{booktitle}{International
  Workshop on Brainlesion: Glioma, Multiple Sclerosis, Stroke and Traumatic
  Brain Injuries}, \bibinfo{organization}{Springer}. pp.
  \bibinfo{pages}{231--243}.
\bibitem[{Coutts et~al.(2003)Coutts, Simon, Tomanek, Barber, Chan, Hudon,
  Mitchell, Frayne, Eliasziw, Buchan et~al.}]{coutts2003reliability}
\bibinfo{author}{Coutts, S.B.}, \bibinfo{author}{Simon, J.E.},
  \bibinfo{author}{Tomanek, A.I.}, \bibinfo{author}{Barber, P.A.},
  \bibinfo{author}{Chan, J.}, \bibinfo{author}{Hudon, M.E.},
  \bibinfo{author}{Mitchell, J.R.}, \bibinfo{author}{Frayne, R.},
  \bibinfo{author}{Eliasziw, M.}, \bibinfo{author}{Buchan, A.M.}, et~al.,
  \bibinfo{year}{2003}.
\newblock \bibinfo{title}{Reliability of assessing percentage of
  diffusion-perfusion mismatch}.
\newblock \bibinfo{journal}{Stroke} \bibinfo{volume}{34},
  \bibinfo{pages}{1681--1683}.
\bibitem[{El~Tawil and Muir(2017)}]{el2017thrombolysis}
\bibinfo{author}{El~Tawil, S.}, \bibinfo{author}{Muir, K.W.},
  \bibinfo{year}{2017}.
\newblock \bibinfo{title}{Thrombolysis and thrombectomy for acute ischaemic
  stroke}.
\newblock \bibinfo{journal}{Clinical Medicine} \bibinfo{volume}{17},
  \bibinfo{pages}{161--165}.
\bibitem[{Gonzalez et~al.(2007)Gonzalez, Hirsch, Koroshetz, Lev and
  Schaefer}]{gonzalez2007acute}
\bibinfo{author}{Gonzalez, R.}, \bibinfo{author}{Hirsch, J.},
  \bibinfo{author}{Koroshetz, W.}, \bibinfo{author}{Lev, M.},
  \bibinfo{author}{Schaefer, P.}, \bibinfo{year}{2007}.
\newblock \bibinfo{title}{Acute ischemic stroke: imaging and intervention}.
\newblock \bibinfo{journal}{{American} Journal of Neuroradiology}
  \bibinfo{volume}{28}, \bibinfo{pages}{1622}.
\bibitem[{Gonz{\'a}lez et~al.(2011)Gonz{\'a}lez, Hirsch, Koroshetz, Lev and
  Schaefer}]{gonzalez2011acute}
\bibinfo{author}{Gonz{\'a}lez, R.G.}, \bibinfo{author}{Hirsch, J.A.},
  \bibinfo{author}{Koroshetz, W.}, \bibinfo{author}{Lev, M.H.},
  \bibinfo{author}{Schaefer, P.W.}, \bibinfo{year}{2011}.
\newblock \bibinfo{title}{Acute ischemic stroke}.
\newblock \bibinfo{publisher}{Springer}.
\bibitem[{Grysiewicz et~al.(2008)Grysiewicz, Thomas and
  Pandey}]{grysiewicz2008epidemiology}
\bibinfo{author}{Grysiewicz, R.A.}, \bibinfo{author}{Thomas, K.},
  \bibinfo{author}{Pandey, D.K.}, \bibinfo{year}{2008}.
\newblock \bibinfo{title}{Epidemiology of ischemic and hemorrhagic stroke:
  incidence, prevalence, mortality, and risk factors}.
\newblock \bibinfo{journal}{Neurologic clinics} \bibinfo{volume}{26},
  \bibinfo{pages}{871--895}.
\bibitem[{Higashida et~al.(2003)Higashida, Furlan, Roberts, Tomsick, Connors,
  Barr, Dillon, Warach, Broderick, Tilley et~al.}]{higashida2003trial}
\bibinfo{author}{Higashida, R.T.}, \bibinfo{author}{Furlan, A.J.},
  \bibinfo{author}{Roberts, H.}, \bibinfo{author}{Tomsick, T.},
  \bibinfo{author}{Connors, B.}, \bibinfo{author}{Barr, J.},
  \bibinfo{author}{Dillon, W.}, \bibinfo{author}{Warach, S.},
  \bibinfo{author}{Broderick, J.}, \bibinfo{author}{Tilley, B.}, et~al.,
  \bibinfo{year}{2003}.
\newblock \bibinfo{title}{Trial design and reporting standards for
  intraarterial cerebral thrombolysis for acute ischemic stroke}.
\newblock \bibinfo{journal}{Journal of Vascular and Interventional Radiology}
  \bibinfo{volume}{14}, \bibinfo{pages}{E1--E31}.
\bibitem[{Hinton(2012)}]{hinton2012practical}
\bibinfo{author}{Hinton, G.E.}, \bibinfo{year}{2012}.
\newblock \bibinfo{title}{A practical guide to training restricted {Boltzmann}
  machines}, in: \bibinfo{booktitle}{Neural networks: Tricks of the trade}.
  \bibinfo{publisher}{Springer}, pp. \bibinfo{pages}{599--619}.
\bibitem[{Hochreiter and Schmidhuber(1997)}]{hochreiter1997long}
\bibinfo{author}{Hochreiter, S.}, \bibinfo{author}{Schmidhuber, J.},
  \bibinfo{year}{1997}.
\newblock \bibinfo{title}{Long short-term memory}.
\newblock \bibinfo{journal}{Neural computation} \bibinfo{volume}{9},
  \bibinfo{pages}{1735--1780}.
\bibitem[{Jenkinson et~al.(2012)Jenkinson, Beckmann, Behrens, Woolrich and
  Smith}]{jenkinson2012fsl}
\bibinfo{author}{Jenkinson, M.}, \bibinfo{author}{Beckmann, C.F.},
  \bibinfo{author}{Behrens, T.E.}, \bibinfo{author}{Woolrich, M.W.},
  \bibinfo{author}{Smith, S.M.}, \bibinfo{year}{2012}.
\newblock \bibinfo{title}{Fsl}.
\newblock \bibinfo{journal}{Neuroimage} \bibinfo{volume}{62},
  \bibinfo{pages}{782--790}.
\bibitem[{Kamnitsas et~al.(2017)Kamnitsas, Ledig, Newcombe, Simpson, Kane,
  Menon, Rueckert and Glocker}]{kamnitsas2017efficient}
\bibinfo{author}{Kamnitsas, K.}, \bibinfo{author}{Ledig, C.},
  \bibinfo{author}{Newcombe, V.F.}, \bibinfo{author}{Simpson, J.P.},
  \bibinfo{author}{Kane, A.D.}, \bibinfo{author}{Menon, D.K.},
  \bibinfo{author}{Rueckert, D.}, \bibinfo{author}{Glocker, B.},
  \bibinfo{year}{2017}.
\newblock \bibinfo{title}{Efficient multi-scale 3d {CNN} with fully connected
  {CRF} for accurate brain lesion segmentation}.
\newblock \bibinfo{journal}{Medical image analysis} \bibinfo{volume}{36},
  \bibinfo{pages}{61--78}.
\bibitem[{Kemmling et~al.(2015)Kemmling, Flottmann, Forkert, Minnerup, Heindel,
  Thomalla, Eckert, Knauth, Psychogios, Langner and Fiehler}]{kemmling2015}
\bibinfo{author}{Kemmling, A.}, \bibinfo{author}{Flottmann, F.},
  \bibinfo{author}{Forkert, N.D.}, \bibinfo{author}{Minnerup, J.},
  \bibinfo{author}{Heindel, W.}, \bibinfo{author}{Thomalla, G.},
  \bibinfo{author}{Eckert, B.}, \bibinfo{author}{Knauth, M.},
  \bibinfo{author}{Psychogios, M.}, \bibinfo{author}{Langner, S.},
  \bibinfo{author}{Fiehler, J.}, \bibinfo{year}{2015}.
\newblock \bibinfo{title}{Multivariate dynamic prediction of ischemic
  infarction and tissue salvage as a function of time and degree of
  recanalization}.
\newblock \bibinfo{journal}{Journal of Cerebral Blood Flow \& Metabolism}
  \bibinfo{volume}{35}, \bibinfo{pages}{1397--1405}.
\bibitem[{Kistler et~al.(2013)Kistler, Bonaretti, Pfahrer, Niklaus and
  B{\"u}chler}]{kistler2013virtual}
\bibinfo{author}{Kistler, M.}, \bibinfo{author}{Bonaretti, S.},
  \bibinfo{author}{Pfahrer, M.}, \bibinfo{author}{Niklaus, R.},
  \bibinfo{author}{B{\"u}chler, P.}, \bibinfo{year}{2013}.
\newblock \bibinfo{title}{The virtual skeleton database: an open access
  repository for biomedical research and collaboration}.
\newblock \bibinfo{journal}{Journal of medical Internet research}
  \bibinfo{volume}{15}.
\bibitem[{Krizhevsky et~al.(2012)Krizhevsky, Sutskever and
  Hinton}]{krizhevsky2012imagenet}
\bibinfo{author}{Krizhevsky, A.}, \bibinfo{author}{Sutskever, I.},
  \bibinfo{author}{Hinton, G.E.}, \bibinfo{year}{2012}.
\newblock \bibinfo{title}{Imagenet classification with deep convolutional
  neural networks}, in: \bibinfo{booktitle}{Advances in neural information
  processing systems}, pp. \bibinfo{pages}{1097--1105}.
\bibitem[{Labeyrie et~al.(2012)Labeyrie, Turc, Hess, Hervo, Mas, Meder, Baron,
  Touz{\'e} and Oppenheim}]{labeyrie2012diffusion}
\bibinfo{author}{Labeyrie, M.A.}, \bibinfo{author}{Turc, G.},
  \bibinfo{author}{Hess, A.}, \bibinfo{author}{Hervo, P.},
  \bibinfo{author}{Mas, J.L.}, \bibinfo{author}{Meder, J.F.},
  \bibinfo{author}{Baron, J.C.}, \bibinfo{author}{Touz{\'e}, E.},
  \bibinfo{author}{Oppenheim, C.}, \bibinfo{year}{2012}.
\newblock \bibinfo{title}{Diffusion lesion reversal after thrombolysis: a mr
  correlate of early neurological improvement}.
\newblock \bibinfo{journal}{Stroke} \bibinfo{volume}{43},
  \bibinfo{pages}{2986--2991}.
\bibitem[{Liebeskind(2003)}]{liebeskind2003collateral}
\bibinfo{author}{Liebeskind, D.S.}, \bibinfo{year}{2003}.
\newblock \bibinfo{title}{Collateral circulation}.
\newblock \bibinfo{journal}{Stroke} \bibinfo{volume}{34},
  \bibinfo{pages}{2279--2284}.
\bibitem[{Lucas and Heinrich(2017)}]{lucas20172d}
\bibinfo{author}{Lucas, C.}, \bibinfo{author}{Heinrich, M.P.},
  \bibinfo{year}{2017}.
\newblock \bibinfo{title}{2d multi-scale res-net for stroke segmentation}.
\bibitem[{Maier et~al.(2017)Maier, Menze, von~der Gablentz, H{\"a}ni, Heinrich,
  Liebrand, Winzeck, Basit, Bentley, Chen et~al.}]{maier2017isles}
\bibinfo{author}{Maier, O.}, \bibinfo{author}{Menze, B.H.},
  \bibinfo{author}{von~der Gablentz, J.}, \bibinfo{author}{H{\"a}ni, L.},
  \bibinfo{author}{Heinrich, M.P.}, \bibinfo{author}{Liebrand, M.},
  \bibinfo{author}{Winzeck, S.}, \bibinfo{author}{Basit, A.},
  \bibinfo{author}{Bentley, P.}, \bibinfo{author}{Chen, L.}, et~al.,
  \bibinfo{year}{2017}.
\newblock \bibinfo{title}{{ISLES} 2015-a public evaluation benchmark for
  ischemic stroke lesion segmentation from multispectral {MRI}}.
\newblock \bibinfo{journal}{Medical image analysis} \bibinfo{volume}{35},
  \bibinfo{pages}{250--269}.
\bibitem[{McKinley et~al.(2017)McKinley, H{\"a}ni, Gralla, El-Koussy, Bauer,
  Arnold, Fischer, Jung, Mattmann, Reyes et~al.}]{mckinley}
\bibinfo{author}{McKinley, R.}, \bibinfo{author}{H{\"a}ni, L.},
  \bibinfo{author}{Gralla, J.}, \bibinfo{author}{El-Koussy, M.},
  \bibinfo{author}{Bauer, S.}, \bibinfo{author}{Arnold, M.},
  \bibinfo{author}{Fischer, U.}, \bibinfo{author}{Jung, S.},
  \bibinfo{author}{Mattmann, K.}, \bibinfo{author}{Reyes, M.}, et~al.,
  \bibinfo{year}{2017}.
\newblock \bibinfo{title}{Fully automated stroke tissue estimation using random
  forest classifiers (faster)}.
\newblock \bibinfo{journal}{Journal of Cerebral Blood Flow \& Metabolism}
  \bibinfo{volume}{37}, \bibinfo{pages}{2728--2741}.
\bibitem[{Memezawa et~al.(1992)Memezawa, Smith and
  Siesj{\"o}}]{memezawa1992penumbral}
\bibinfo{author}{Memezawa, H.}, \bibinfo{author}{Smith, M.L.},
  \bibinfo{author}{Siesj{\"o}, B.K.}, \bibinfo{year}{1992}.
\newblock \bibinfo{title}{Penumbral tissues salvaged by reperfusion following
  middle cerebral artery occlusion in rats.}
\newblock \bibinfo{journal}{Stroke} \bibinfo{volume}{23},
  \bibinfo{pages}{552--559}.
\bibitem[{Menon et~al.(2015)Menon, Qazi, Nambiar, Foster, Yeatts, Liebeskind,
  Jovin, Goyal, Hill, Tomsick et~al.}]{menon2015differential}
\bibinfo{author}{Menon, B.K.}, \bibinfo{author}{Qazi, E.},
  \bibinfo{author}{Nambiar, V.}, \bibinfo{author}{Foster, L.D.},
  \bibinfo{author}{Yeatts, S.D.}, \bibinfo{author}{Liebeskind, D.},
  \bibinfo{author}{Jovin, T.G.}, \bibinfo{author}{Goyal, M.},
  \bibinfo{author}{Hill, M.D.}, \bibinfo{author}{Tomsick, T.A.}, et~al.,
  \bibinfo{year}{2015}.
\newblock \bibinfo{title}{Differential effect of baseline computed tomographic
  angiography collaterals on clinical outcome in patients enrolled in the
  interventional management of stroke iii trial}.
\newblock \bibinfo{journal}{Stroke} \bibinfo{volume}{46},
  \bibinfo{pages}{1239--1244}.
\bibitem[{Milletari et~al.(2016)Milletari, Navab and Ahmadi}]{milletari2016v}
\bibinfo{author}{Milletari, F.}, \bibinfo{author}{Navab, N.},
  \bibinfo{author}{Ahmadi, S.A.}, \bibinfo{year}{2016}.
\newblock \bibinfo{title}{V-net: Fully convolutional neural networks for
  volumetric medical image segmentation}, in: \bibinfo{booktitle}{3D Vision
  (3DV), 2016 Fourth International Conference on},
  \bibinfo{organization}{IEEE}. pp. \bibinfo{pages}{565--571}.
\bibitem[{Mok and Chung(2017)}]{mokdeep}
\bibinfo{author}{Mok, T.C.}, \bibinfo{author}{Chung, A.C.},
  \bibinfo{year}{2017}.
\newblock \bibinfo{title}{Deep adversarial networks for stroke lesion
  segmentation} .
\bibitem[{Monteiro and Oliveira(2017)}]{monteirofully}
\bibinfo{author}{Monteiro, M.}, \bibinfo{author}{Oliveira, A.L.},
  \bibinfo{year}{2017}.
\newblock \bibinfo{title}{Fully convolutional neural network for 3d stroke
  lesion segmentation} .
\bibitem[{Nair and Hinton(2010)}]{nair2010rectified}
\bibinfo{author}{Nair, V.}, \bibinfo{author}{Hinton, G.E.},
  \bibinfo{year}{2010}.
\newblock \bibinfo{title}{Rectified linear units improve restricted {Boltzmann}
  machines}, in: \bibinfo{booktitle}{Proceedings of the 27th international
  conference on machine learning (ICML-10)}, pp. \bibinfo{pages}{807--814}.
\bibitem[{Nielsen et~al.(2018)Nielsen, Hansen, Tietze and
  Mouridsen}]{nielsen2018prediction}
\bibinfo{author}{Nielsen, A.}, \bibinfo{author}{Hansen, M.B.},
  \bibinfo{author}{Tietze, A.}, \bibinfo{author}{Mouridsen, K.},
  \bibinfo{year}{2018}.
\newblock \bibinfo{title}{Prediction of tissue outcome and assessment of
  treatment effect in acute ischemic stroke using deep learning}.
\newblock \bibinfo{journal}{Stroke} \bibinfo{volume}{49},
  \bibinfo{pages}{1394--1401}.
\bibitem[{Niu et~al.(2018)Niu, Gong, Xu, Pauly and Zaharchuk}]{niu2018improved}
\bibinfo{author}{Niu, Y.}, \bibinfo{author}{Gong, E.}, \bibinfo{author}{Xu,
  J.}, \bibinfo{author}{Pauly, J.}, \bibinfo{author}{Zaharchuk, G.},
  \bibinfo{year}{2018}.
\newblock \bibinfo{title}{Improved prediction of the final infarct from acute
  stroke neuroimaging using deep learning}, in: \bibinfo{booktitle}{STROKE}.
\bibitem[{Oliveira et~al.(2018)Oliveira, Pereira and
  Silva}]{oliveira2018retinal}
\bibinfo{author}{Oliveira, A.}, \bibinfo{author}{Pereira, S.},
  \bibinfo{author}{Silva, C.A.}, \bibinfo{year}{2018}.
\newblock \bibinfo{title}{Retinal vessel segmentation based on fully
  convolutional neural networks}.
\newblock \bibinfo{journal}{Expert Systems with Applications}
  \bibinfo{volume}{112}, \bibinfo{pages}{229--242}.
\bibitem[{Pereira et~al.(2018)Pereira, Meier, McKinley, Wiest, Alves, Silva and
  Reyes}]{pereira2018enhancing}
\bibinfo{author}{Pereira, S.}, \bibinfo{author}{Meier, R.},
  \bibinfo{author}{McKinley, R.}, \bibinfo{author}{Wiest, R.},
  \bibinfo{author}{Alves, V.}, \bibinfo{author}{Silva, C.A.},
  \bibinfo{author}{Reyes, M.}, \bibinfo{year}{2018}.
\newblock \bibinfo{title}{Enhancing interpretability of automatically extracted
  machine learning features: application to a {RBM-Random Forest} system on
  brain lesion segmentation}.
\newblock \bibinfo{journal}{Medical image analysis} \bibinfo{volume}{44},
  \bibinfo{pages}{228--244}.
\bibitem[{{Pereira} et~al.(2019){Pereira}, {Pinto}, {Amorim}, {Ribeiro},
  {Alves} and {Silva}}]{pereira2018adaptive}
\bibinfo{author}{{Pereira}, S.}, \bibinfo{author}{{Pinto}, A.},
  \bibinfo{author}{{Amorim}, J.}, \bibinfo{author}{{Ribeiro}, A.},
  \bibinfo{author}{{Alves}, V.}, \bibinfo{author}{{Silva}, C.A.},
  \bibinfo{year}{2019}.
\newblock \bibinfo{title}{Adaptive feature recombination and recalibration for
  semantic segmentation with fully convolutional networks}.
\newblock \bibinfo{journal}{IEEE Transactions on Medical Imaging}
  \bibinfo{volume}{38}, \bibinfo{pages}{2914--2925}.
\newblock \DOIprefix\doi{10.1109/TMI.2019.2918096}.
\bibitem[{Pinto et~al.(2018a)Pinto, McKinley, Alves, Wiest, Silva, Reyes
  et~al.}]{pinto2018stroke}
\bibinfo{author}{Pinto, A.}, \bibinfo{author}{McKinley, R.},
  \bibinfo{author}{Alves, V.}, \bibinfo{author}{Wiest, R.},
  \bibinfo{author}{Silva, C.A.}, \bibinfo{author}{Reyes, M.}, et~al.,
  \bibinfo{year}{2018}a.
\newblock \bibinfo{title}{Stroke lesion outcome prediction based on {MRI}
  imaging combined with clinical information}.
\newblock \bibinfo{journal}{Frontiers in Neurology} \bibinfo{volume}{9},
  \bibinfo{pages}{1060}.
\bibitem[{Pinto et~al.(2018b)Pinto, Pereira, Meier, Alves, Wiest, Silva and
  Reyes}]{pinto2018enhancing}
\bibinfo{author}{Pinto, A.}, \bibinfo{author}{Pereira, S.},
  \bibinfo{author}{Meier, R.}, \bibinfo{author}{Alves, V.},
  \bibinfo{author}{Wiest, R.}, \bibinfo{author}{Silva, C.A.},
  \bibinfo{author}{Reyes, M.}, \bibinfo{year}{2018}b.
\newblock \bibinfo{title}{Enhancing clinical {MRI} perfusion maps with
  data-driven maps of complementary nature for lesion outcome prediction}, in:
  \bibinfo{booktitle}{Medical Image Computing and Computer Assisted
  Intervention -- MICCAI 2018}, pp. \bibinfo{pages}{107--115}.
\bibitem[{Pisov et~al.(2017)Pisov, Belyaev and Krivov}]{pisovneural}
\bibinfo{author}{Pisov, M.}, \bibinfo{author}{Belyaev, M.},
  \bibinfo{author}{Krivov, E.}, \bibinfo{year}{2017}.
\newblock \bibinfo{title}{Neural networks ensembles for ischemic stroke lesion
  segmentation} .
\bibitem[{Powers et~al.(2018)Powers, Rabinstein, Ackerson, Adeoye, Bambakidis,
  Becker, Biller, Brown, Demaerschalk, Hoh et~al.}]{powers20182018}
\bibinfo{author}{Powers, W.J.}, \bibinfo{author}{Rabinstein, A.A.},
  \bibinfo{author}{Ackerson, T.}, \bibinfo{author}{Adeoye, O.M.},
  \bibinfo{author}{Bambakidis, N.C.}, \bibinfo{author}{Becker, K.},
  \bibinfo{author}{Biller, J.}, \bibinfo{author}{Brown, M.},
  \bibinfo{author}{Demaerschalk, B.M.}, \bibinfo{author}{Hoh, B.}, et~al.,
  \bibinfo{year}{2018}.
\newblock \bibinfo{title}{2018 guidelines for the early management of patients
  with acute ischemic stroke: a guideline for healthcare professionals from the
  {American Heart Association/American Stroke Association}}.
\newblock \bibinfo{journal}{Stroke} \bibinfo{volume}{49},
  \bibinfo{pages}{e46--e99}.
\bibitem[{Rekik et~al.(2012)Rekik, Allassonni{\`e}re, Carpenter and
  Wardlaw}]{rekik2012medical}
\bibinfo{author}{Rekik, I.}, \bibinfo{author}{Allassonni{\`e}re, S.},
  \bibinfo{author}{Carpenter, T.K.}, \bibinfo{author}{Wardlaw, J.M.},
  \bibinfo{year}{2012}.
\newblock \bibinfo{title}{Medical image analysis methods in {MR/CT}-imaged
  acute-subacute ischemic stroke lesion: Segmentation, prediction and insights
  into dynamic evolution simulation models. a critical appraisal}.
\newblock \bibinfo{journal}{NeuroImage: Clinical} \bibinfo{volume}{1},
  \bibinfo{pages}{164--178}.
\bibitem[{Robben et~al.(2020)Robben, Boers, Marquering, Langezaal, Roos, van
  Oostenbrugge, van Zwam, Dippel, Majoie, van~der Lugt
  et~al.}]{robben2020prediction}
\bibinfo{author}{Robben, D.}, \bibinfo{author}{Boers, A.M.},
  \bibinfo{author}{Marquering, H.A.}, \bibinfo{author}{Langezaal, L.L.},
  \bibinfo{author}{Roos, Y.B.}, \bibinfo{author}{van Oostenbrugge, R.J.},
  \bibinfo{author}{van Zwam, W.H.}, \bibinfo{author}{Dippel, D.W.},
  \bibinfo{author}{Majoie, C.B.}, \bibinfo{author}{van~der Lugt, A.}, et~al.,
  \bibinfo{year}{2020}.
\newblock \bibinfo{title}{Prediction of final infarct volume from native ct
  perfusion and treatment parameters using deep learning}.
\newblock \bibinfo{journal}{Medical image analysis} \bibinfo{volume}{59},
  \bibinfo{pages}{101589}.
\bibitem[{Robben and Suetens(2017)}]{robben2017dual}
\bibinfo{author}{Robben, D.}, \bibinfo{author}{Suetens, P.},
  \bibinfo{year}{2017}.
\newblock \bibinfo{title}{Dual-scale fully convolutional neural network for
  final infarct prediction}, in: \bibinfo{booktitle}{Ischemic stroke lesion
  segmentation-ISLES challenge 2017, held in conjunction with MICCAI 2017,
  Date: 2017/09/10-2017/09/10, Location: Quebec City, Quebec, Canada}.
\bibitem[{Ronneberger et~al.(2015)Ronneberger, Fischer and
  Brox}]{ronneberger2015u}
\bibinfo{author}{Ronneberger, O.}, \bibinfo{author}{Fischer, P.},
  \bibinfo{author}{Brox, T.}, \bibinfo{year}{2015}.
\newblock \bibinfo{title}{U-net: Convolutional networks for biomedical image
  segmentation}, in: \bibinfo{booktitle}{International Conference on Medical
  image computing and computer-assisted intervention},
  \bibinfo{organization}{Springer}. pp. \bibinfo{pages}{234--241}.
\bibitem[{Rose et~al.(2001)Rose, Chalk, Griffin, Janke, Chen, McLachan, Peel,
  Zelaya, Markus, Jones et~al.}]{rose2001mri}
\bibinfo{author}{Rose, S.E.}, \bibinfo{author}{Chalk, J.B.},
  \bibinfo{author}{Griffin, M.P.}, \bibinfo{author}{Janke, A.L.},
  \bibinfo{author}{Chen, F.}, \bibinfo{author}{McLachan, G.J.},
  \bibinfo{author}{Peel, D.}, \bibinfo{author}{Zelaya, F.O.},
  \bibinfo{author}{Markus, H.S.}, \bibinfo{author}{Jones, D.K.}, et~al.,
  \bibinfo{year}{2001}.
\newblock \bibinfo{title}{{MRI} based diffusion and perfusion predictive model
  to estimate stroke evolution}.
\newblock \bibinfo{journal}{Magnetic resonance imaging} \bibinfo{volume}{19},
  \bibinfo{pages}{1043--1053}.
\bibitem[{Rumelhart and McClelland(1986)}]{rumelhart1986parallel}
\bibinfo{author}{Rumelhart, D.E.}, \bibinfo{author}{McClelland, J.L.},
  \bibinfo{year}{1986}.
\newblock \bibinfo{title}{Parallel distributed processing: explorations in the
  microstructure of cognition. volume 1. foundations} .
\bibitem[{Scalzo et~al.(2012)Scalzo, Hao, Alger, Hu and
  Liebeskind}]{scalzo2012}
\bibinfo{author}{Scalzo, F.}, \bibinfo{author}{Hao, Q.},
  \bibinfo{author}{Alger, J.R.}, \bibinfo{author}{Hu, X.},
  \bibinfo{author}{Liebeskind, D.S.}, \bibinfo{year}{2012}.
\newblock \bibinfo{title}{Regional prediction of tissue fate in acute ischemic
  stroke}.
\newblock \bibinfo{journal}{Annals of Biomedical Engineering}
  \bibinfo{volume}{40}, \bibinfo{pages}{2177--2187}.
\bibitem[{{SMIR Online Platform}(2017)}]{Site}
\bibinfo{author}{{SMIR Online Platform}}, \bibinfo{year}{2017}.
\newblock \bibinfo{title}{Ischemic stroke lesion segmentation 2017}.
\newblock \bibinfo{howpublished}{\url{https://www.smir.ch/ISLES/Start2017}}.
\newblock \bibinfo{note}{[Acessed: 2019-11-18]}.
\bibitem[{Vinh et~al.(2010)Vinh, Epps and Bailey}]{vinh2010information}
\bibinfo{author}{Vinh, N.X.}, \bibinfo{author}{Epps, J.},
  \bibinfo{author}{Bailey, J.}, \bibinfo{year}{2010}.
\newblock \bibinfo{title}{Information theoretic measures for clusterings
  comparison: Variants, properties, normalization and correction for chance}.
\newblock \bibinfo{journal}{Journal of Machine Learning Research}
  \bibinfo{volume}{11}, \bibinfo{pages}{2837--2854}.
\bibitem[{Visin et~al.(2016)Visin, Ciccone, Romero, Kastner, Cho, Bengio,
  Matteucci and Courville}]{visin2016reseg}
\bibinfo{author}{Visin, F.}, \bibinfo{author}{Ciccone, M.},
  \bibinfo{author}{Romero, A.}, \bibinfo{author}{Kastner, K.},
  \bibinfo{author}{Cho, K.}, \bibinfo{author}{Bengio, Y.},
  \bibinfo{author}{Matteucci, M.}, \bibinfo{author}{Courville, A.},
  \bibinfo{year}{2016}.
\newblock \bibinfo{title}{Reseg: A recurrent neural network-based model for
  semantic segmentation}, in: \bibinfo{booktitle}{Proceedings of the IEEE
  Conference on Computer Vision and Pattern Recognition Workshops}, pp.
  \bibinfo{pages}{41--48}.
\bibitem[{Winzeck et~al.(2018)Winzeck, Hakim, McKinley, Pinto, Alves, Silva,
  Pisov, Krivov, Belyaev, Monteiro et~al.}]{winzeck2018isles}
\bibinfo{author}{Winzeck, S.}, \bibinfo{author}{Hakim, A.},
  \bibinfo{author}{McKinley, R.}, \bibinfo{author}{Pinto, J.A.A.D.S.},
  \bibinfo{author}{Alves, V.}, \bibinfo{author}{Silva, C.},
  \bibinfo{author}{Pisov, M.}, \bibinfo{author}{Krivov, E.},
  \bibinfo{author}{Belyaev, M.}, \bibinfo{author}{Monteiro, M.}, et~al.,
  \bibinfo{year}{2018}.
\newblock \bibinfo{title}{{ISLES} 2016 \& 2017-benchmarking ischemic stroke
  lesion outcome prediction based on multispectral {MRI}}.
\newblock \bibinfo{journal}{Frontiers in Neurology} \bibinfo{volume}{9},
  \bibinfo{pages}{679}.
\bibitem[{{World Health Organization} et~al.(2014)}]{world2014global}
\bibinfo{author}{{World Health Organization}}, et~al., \bibinfo{year}{2014}.
\newblock \bibinfo{title}{Global status report on noncommunicable diseases
  2014.}
\newblock \bibinfo{publisher}{World Health Organization}.
\bibitem[{Zeiler and Fergus(2014)}]{zeiler2014visualizing}
\bibinfo{author}{Zeiler, M.D.}, \bibinfo{author}{Fergus, R.},
  \bibinfo{year}{2014}.
\newblock \bibinfo{title}{Visualizing and understanding convolutional
  networks}, in: \bibinfo{booktitle}{European conference on computer vision},
  \bibinfo{organization}{Springer}. pp. \bibinfo{pages}{818--833}.
\bibitem[{Zivelonghi and Tamburin(2018)}]{zivelonghi2018mechanical}
\bibinfo{author}{Zivelonghi, C.}, \bibinfo{author}{Tamburin, S.},
  \bibinfo{year}{2018}.
\newblock \bibinfo{title}{Mechanical thrombectomy for acute ischemic stroke:
  the therapeutic window is larger but still “time is brain”}.
\newblock \bibinfo{journal}{Functional neurology} \bibinfo{volume}{33},
  \bibinfo{pages}{5}.

\end{thebibliography}
\end{document}